\documentclass[11pt]{article}
\usepackage[utf8]{inputenc}
\usepackage{bbm,amssymb,amsmath,amsfonts,eurosym,geometry,ulem,graphicx,color,setspace,sectsty,comment,footmisc,natbib,pdflscape,array,pdflscape,caption,subcaption,multirow,tablefootnote,wasysym, changepage, xcolor, threeparttable, booktabs, adjustbox}
\usepackage[hidelinks]{hyperref}
\usepackage{endfloat}
\begin{document}

%--------------------------------------------------------
\begin{titlepage}
\title{Local Labor Market Effects of Mergers and Acquisitions in Developing Countries: Evidence from Brazil}
\author{Vítor Costa\thanks{Economics PhD Candidate at Cornell University. }}
\date{\today}
\maketitle

\begin{abstract}

\noindent 
I use matched employer-employee records merged with corporate tax information from 2003 to 2017 to estimate labor market-wide effects of mergers and acquisitions in Brazil. Labor markets are defined by pairs of commuting zone and industry sector. In the following year of a merger, market size falls by 10.8\%. The employment adjustment is concentrated in merging firms. For the firms not involved in M\&As, I estimate a 1.07\% decline in workers' earnings and a positive, although not significant, increase in their size. Most mergers have a predicted impact of zero points in concentration, measured by the Herfindahl–Hirschman Index (HHI). I spillover firms, earnings decline similarly for mergers with high and low predicted changes in HHI. Contrary to the recent literature on market concentration in developed economies, I find no evidence of oligopsonistic behavior in Brazilian labor markets. 

\vspace{.2in}


\vspace{.2in}

\bigskip
\end{abstract}
\setcounter{page}{0}
\thispagestyle{empty}
\end{titlepage}

%--------------------------------------------------------
\section{Introduction}

A growing body of research in recent years has pointed in the direction of non-competitive behavior in labor markets \citep{card_who_2022}. The formalization of a model where profit-maximizing firms internalize an upward-sloping labor supply curve and, as a result, choose to hire fewer workers at a lower wage rate, dates back to the 1930s \citep{robinson_economics_1933}, but a renewed interest in the subject was sparked by the debate around the reasons for the labor share decline observed in developed economies \citep{autor_fall_2020}. The rise of firms that concentrate ever larger shares of industry sales in a globalized market, the so-called \emph{superstar} firms, has launched many in the profession into the empirical investigation of labor market imperfection related to the increase in the size of employers. While the debate about the underlying causes of the fall in the labor share of the GDP is far from settled \citep{grossman_elusive_2021}, one cannot neglect the burgeoning literature pointing to a negative relationship between labor market outcomes and employment concentration, and proposals for policy intervention to protect workers abound, especially within the reach of antitrust regulation \citep{naidu_antitrust_2018, marinescu_anticompetitive_2019, marinescu_why_2020}.    

Naturally, mergers and acquisitions (M\&As) raise immediate concern about the competitiveness in labor markets. By means of the consolidation of different employers under the same ownership and management, mergers mechanically alter the number of firms competing for labor services and, therefore, might in principle, tilt the balance of bargaining power in an unfavorabe manner to workers. The relationship between a lower number of employers and worse workers' outcomes relies on more than just intuition, and it has both theoretical and empirical grounds. Building upon the monopsony model from \cite{robinson_economics_1933}, \cite{boal_monopsony_1997} show that, similar to the case of oligopoly, a model of \emph{oligopsony} à la Cournot generates wage markdowns that decrease as the number of employers falls, i.e., wages represent a lower fraction of the marginal revenue product of labor as employment gets concentrated among fewer firms. Negative wage elasticities with respect to employment concentration have been confirmed in various contexts \footnote{See \cite{martins_making_2018, lipsius_labor_2018, azar_measuring_2019, marinescu_wages_2021, azar_concentration_2020,  devereux_local_2023, dodini_monopsony_2023, benmelech_strong_2022, rinz_labor_2020, bassanini_labor_2023}.}. Typically, employment concentration is measured by the Herfindahl–Hirschman Index (HHI) over firms' workforce shares in local markets, sometimes defined by a combination of geographical region and industry sector, or region and occupational codes. This measurement is made possible by the use of linked employer-employee administrative datasets.     

In this paper, I investigate the local labor market effects of mergers and acquisitions, and the role played by labor market concentration as a channel of these effects. I combine two different administrative datasets from Brazil that allow me to identify merged and acquired establishments, delineate local labor markets, and causally estimate changes in employment, workers' earnings, and local concentration measured by the HHI. The findings consist of three sequential steps. First, I show that null effects of the M\&As on workers earnings and market HHI cannot be rejected, while employment significantly declines in markets that witness a firm consolidation event. Next, I split the estimation procedure between two separate groups of firms: the ones that participate in M\&As, and the bystander employers in the same market, which I will call spillover firms. The separation of the two types of firms shows that they respond differently to merger activity. The negative employment effects found at the market level are primarily carried out by merging firms, while spillover firms show a small, albeit not significant, increase in their size. By looking at the trajectory of hires and separations, I find that the negative adjustment in merging firms' employment is given by an abrupt decline in new hiring while separations are kept at pre-merger levels for at least one year after the event. 

The third part of the analysis contains the main result of this paper. In order to evaluate if larger increments in local labor market concentration deepen the earnings effects of mergers, as predicted by the oligopsony theory, I compare the estimates in spillover firms between mergers with no change in concentration and mergers at the top of the distribution of concentration shocks. Contrary to the previous literature, I find that the earnings in spillover firms are similar in both cases, which indicates that concentration plays little to no role in explaining the market-wide effects of M\&As in Brazil. In mergers that induce no change in concentration, workers' earnings in spillover firms fall by 1.1\%; the same estimate is obtained from within-market mergers with significant increases in local HHI. Moreover, the events with no change in concentration seem to induce a growth in the size of spillover firms, consistent with the logic that the negative employment adjustment promoted within merging firms prompts an increase in the supply of labor to other firms in the same labor market. This increase in the labor supply available to spillover firms is accommodated by an increase in their employment but at lower wage rate. This is confirmed when I explicitly compare the \emph{pre} and \emph{post}-merger earnings of new hires in spillover firms. 

The empirical strategy of this paper relies on a study event design based on the  comparison of different local labor markets regarding the year that they witness their first merger event. Due to the concern of treatment rollout and heterogeneity giving rise to problematic estimates \citep{sun_estimating_2021, roth_whats_2022, de_chaisemartin_two-way_2022}, I depart from the more traditional two-way fixed effects model and implement the estimation proposed in \cite{callaway_difference--differences_2021}, where the control group to treated markets consists of other markets that will eventually be treated in subsequent years. Also, given that I'm interested in market-wide competition for labor services, I attempt to mitigate the effects of alternative mechanisms connecting merger activity to labor market outcomes. Employment level and wages may be altered by M\&A activity through mechanisms other than the competition for labor services. When two firms in the same industry sector merge, higher price setting in their output market has ramifications for the market for inputs not necessarily related to higher bargaining power with upstream service providers – for instance, lower wages and employment could be driven by the monopolist's decision to produce below the competitive benchmark, absent of changes to their ability to set wages. Another possibility is that merged or acquired firms are better equipped to change the composition of their workforce – in case bigger firms can hire younger, less experienced workers, or workers with lower educational attainment, without hindering productivity, lower observed wages could be merely a consequence of lower compensation corresponding to these attributes. In order to best control for the product market concentration and labor compositional mechanisms, I restrict the analysis to tradable sectors only and, leveraging on the demographic details available in the data, I estimate effects on an earnings measure that takes observable attributes of the workers into account.        

This paper contributes to two strands of the literature on the effects of employer consolidation on labor outcomes. The first one is related to the direct effects of mergers and acquisitions on employment and earnings in acquired and merged establishments \citep{shleifer_breach_1987, brown_impact_1987,conyon_hostile_2001, conyon_impact_2002, lehto_analysing_2008, siegel_assessing_2010, he_mergers_2019, lagaras_mas_2023, guanziroli_does_2022}. Expanding on this literature, I focus not only on target establishments or firms, but I also keep track of earnings and employment at the other employers operating in the same labor market. This observation increases the understanding of market-wide effects of merger activity. The second related literature is the one that directly studies the relationship between labor market concentration and wages, and employment \citep{martins_making_2018, lipsius_labor_2018, azar_measuring_2019, marinescu_wages_2021, azar_concentration_2020,  devereux_local_2023, dodini_monopsony_2023, benmelech_strong_2022, rinz_labor_2020, bassanini_labor_2023}. Here, my contribution is to expand the evidence beyond the context of developed economies, and look at the role of concentration in earnings and employment effects in a developing economy. To the best of my knowledge, this is the first study exploring market-wide effects of merger activity from multiple M\&A events, in a wide range of industry sectors, in a developing economy. The most closely related works to this paper are \cite{arnold_mergers_2022} and \cite{prager_employer_2021}; they both estimate M\&A effects in U.S. labor markets, and study the role of employment concentration in mediating these effects, with the difference that Arnold uses administrative data for a wide range of industry sectors, while Prager and Schmitt focus on the hospital sector. They both confirm that, as predicted by oligopsony theory, mergers that induce little to no effect in concentration also have negligible earnings effects, and significant wage declines are only found among merger events in the top of the distribution of concentration changes. My work departs from theirs in the finding that M\&As with larger concentration increases do not generate sharper wage declines in the Brazilian case, and mergers with predicted null change in concentration are followed by an increase in the size of spillover firms\footnote{While \cite{arnold_mergers_2022} does not report market wide employment effects, employment pre-trends in \cite{prager_employer_2021}'s event studies preclude them from making assertive claims on the size of labor markets after hospital consolidation events.}.   

The remainder of the paper is organized as follows. In Section \ref{sec:data}, I describe the data used in the paper. The main results are reported in Section \ref{sec:results}, where I subdivide the analysis starting from merger activity effects in general, then across different firms in the labor market with respect to their participation in M\&As, and, lastly, across events with different predicted impacts in local concentration. I present the robustness of the findings with respect to the possibility of treatment anticipation in Section \ref{sec:rob}. Section \ref{sec:discuss} offers a discussion of the results in view of the literature and auxiliary evidence of management practices in middle to low-income countries. Section \ref{sec:conc} concludes the paper.     

%--------------------------------------------------------
\section{Data} \label{sec:data}
The data used in this paper is composed of two different administrative releases by the Brazilian federal government. The first one is the \emph{Relação Anual de Informações Sociais} - RAIS, the main source for worker and job characteristics information. The second is \emph{Dados Públicos CNPJ} - DPC, a business registry from which I extract the list of establishments that went through a merger or acquisition. More detail about each dataset is provided below. For a complete step-by-step o the data handling and construction, see Appendix \ref{app:data_mani}.  

\subsection{Worker data - RAIS} \label{RAIS}

RAIS is a matched employer-employee administrative record provided by the Brazilian Ministry of Labor on a yearly basis, and it covers the entirety of the country's formal labor market, which employs around 70\% of the workers \citep{hallak_neto_setor_2012}. In terms of the U.S. Census database, RAIS is similar to the Longitudinal Employer-Household Dynamics - LEHD. 

Each observation of RAIS contains separate worker and establishment identifiers. The establishment identifier is hierarchical, and from its first 8 digits, or its \emph{root}, I can also retrieve the firm identifier. As for the workers, the available characteristics are color/race, sex, age, and educational achievement. In the case of establishments, it is possible to see the city where they are located, and the establishment's 5-digit industry sector. The data also contains variables related to the job itself, such as the average monthly earnings, contractual weekly hours, date of admission, and date of separation, in case the job was terminated within that year. Differently from the U.S. Census LEHD, RAIS also reports the occupation of the worker in that particular job. 

I use the variables of location and industry sector to delineate local labor markets. Local labor market details are discussed in Section \ref{sec:llms}. For this paper, I use RAIS in years ranging from 2002 up to 2017, totaling more than 1.1 billion observations. Access to the data at the individual level is restricted by a confidentiality agreement. 

\subsection{Business Data - DPC}

The \emph{Dados Públicos CNPJ} - DPC, is a business registry made available by \emph{Receita Federal}, the Brazilian tax collection agency. A similar counterpart to the DPC in the U.S. Census system is the Longitudinal Business Database—LBD. The DPC contains information on the universe of establishments ever registered with the agency, and it is updated on a monthly basis. There is a total of more than 42 million observations. Access to the DPC is public, and the files can be downloaded from the revenue agency's website\footnote{See \url{https://www.gov.br/receitafederal/pt-br/assuntos/orientacao-tributaria/cadastros/consultas/dados-publicos-cnpj}.} 

From the DPC, it is possible to see an establishment's identifier, postal code, industry sector, and, to some extent, its capital table. Primarily important for this paper, DPC also discloses the variable describing the reason for the termination of a business. By law, anytime a firm or establishment is acquired by or merged with another, its identifier is retired, and a new one is issued for the newly created enterprise. Therefore, merged and acquired establishments can be flagged by the reported reason its identifier was retired. Other grounds for business termination include bankruptcy and various forms of tax penalties. 

Despite the richness of details in the business registry, one cannot use it to identify all the parties involved in an M\&A event. Only the \emph{acquired} or \emph{pre-merger} entities are flagged as having been targeted by an M\&A. However, the identifiers of the firms and establishments after the consolidation are not reported, i.e., the \emph{acquirer} and newly-\emph{merged} firm identifiers are not available. The identification of all parties is important to distinguish mergers that happen within any given labor market from those between employers previously operating in separate markets – such distinction will be used to derive an important result regarding the role of employment concentration and the outcome effects of M\&As. In order to identify all firms involved in a concentration event, I use the dynamic feature of the worker data, RAIS. More specifically, I observe the flows of workers departing from acquired establishments to help me identify the acquirer firm. The most common destination of these workers in the last year I observe acquired establishments is considered to be the acquirer business\footnote{Figure \ref{fig:top_dest}, in Appendix \ref{app:top_dest}, shows the distribution of coalition sizes leaving acquired establishments in years preceding the M\&A.}.

\subsection{Commuting Zones}
The commuting zones are imported from the Brazilian Decennial Census of 2010. Each commuting zone represents contiguous cities between which people commute either for study or work. The worker data is merged with the commuting zone information using the city codes in both datasets. 

%--------------------------------------------------------
\section{Empirical Strategy} \label{sec:model}

The parameter of interest in this paper is the average treatment effect of an M\&A on local labor markets. A local labor market is defined as pairs of commuting zone and industry sector code. The main outcome variables are the workers' earnings, employment, and the local level of employment concentration. This section presents more detail about the estimation procedure and the identification assumptions. 

%--------------------------------------------------------

\subsection{\label{sec:llms} Local Labor Markets}

In order to measure the employment concentration of labor markets, one has to choose how to define the labor markets in the first place. Recent studies have used either a combination of industry sector codes and commuting zone \citep{rinz_labor_2020, benmelech_strong_2022, prager_employer_2021, arnold_mergers_2022}, occupation codes and commuting zone \citep{azar_measuring_2019, azar_concentration_2020, marinescu_wages_2021}, or a data-driven approach that leverages on worker-flow observations \citep{nimczik_job_2018}. In this paper, I will use the definition of local labor markets based on industry codes and commuting zones, i.e., each market will represent an industry activity geographically located in economically integrated cities. One drawback of this definition is that not all workers in a labor market are equally bound to the same industry sector or commuting zone. Accountants, for instance, have skills easily transferrable across different industries within the same region, while nurses are bound by job opportunities in the health sector only, and thus will likely be more willing to switch cities if necessary. On the other hand, I see two advantages to this approach – (i) this is a definition readily available to researchers and policymakers alike, as it does not require further modeling assumptions based on other characteristics of jobs or workers, and (ii) it makes the present investigation comparable to most other studies in the related literature using a similar definition of labor markets.      

As I am interested in the labor market effects of mergers, I will also restrict the analysis to  tradable sectors only. The idea is that unobserved changes in \emph{product} markets that possibly follow M\&As might explain some, if not all, of the post-merger wage and employment decline in treated labor markets. By virtue of increased pricing power on the product market after an M\&A, bigger consolidated firms may decrease their output level and, therefore, reduce their demand for labor. Any declines in wages and employment could then be, to some extent, the result of lower overall demand for labor followed by increased monopoly power, and not necessarily a consequence of any changes in the competition for labor services. A strategy to deal with changes in downstream competition is to narrow the analysis to tradable sectors only \citep{arnold_mergers_2022}. The rationale behind this strategy is that the competition of foreign products and services precludes the rise in the market power of the merged firms, thus shutting down the monopoly channel connecting M\&A events to labor market outcomes\footnote{The selection of tradable industries follows the classification in \cite{dix-carneiro_trade_2017}.}.

The employment concentration measure I will use in the remainder of the paper is the Herfindahl–Hirschman Index (HHI) computed at the local labor market level in each year. More specifically, for a market $m$ and its set of firms $\mathcal{F}_m$, the HHI in year $t$ is given by

$$ HHI_{m,t} = \sum_{f\in \mathcal{F}_m} s_{f,t}^2$$

\noindent
where $s_{f,t}$ is the percentage share of employment of firm $f \in \mathcal{F}_m$ as of December 31st in year $t$. When the local labor market has one single employer, the HHI registers 10,000 points. For an HHI level of $h$ points, the equivalent number $n$ of equally sized employers is given by $n=10,000/h$. Also, if two employers with shares $s_1$ and $s_2$ merge, the equivalent increase in HHI, absent of employment effects, is given by the product $2s_1s_2$. Lastly, an increase of $\delta$ points in HHI is equivalent to the merger of two equally sized firms with employment shares equal to $\sqrt{\delta/2}$ each.    

\subsection{Event Study Setup}

It is possible to assume that consolidation of employers might have lingering effects on local labor markets, and therefore the measure of effects across different lengths of time after mergers take place should be part of the analysis. More so, for the years before the merger, it is also important to observe the difference between the outcomes of treated markets and their pool of control units. Thus, I will base the empirical analysis on the estimation of an event study around the time that local labor markets experience a merger. 

The predominant approach to specifying the event study with multiple cross-sectional units is to use a two-way fixed effects equation with leads and lags of an indicator variable for the treated units \citep{callaway_difference--differences_2021}. However, some concerns raised in the recent literature investigating the two-way fixed effect approach seem applicable in the context of this paper \citep{roth_whats_2022}. First, there is variation in the treatment timing of local labor markets, i.e., not all markets witness a merger in the same year. Second, the treatment effects may not be the same for different cohorts of treated markets, e.g., the treatment effect on markets hosting a merger in 2008 may differ from that of markets treated in 2010, and so on. With multiple treatment periods and heterogeneous effects across cohorts, the two-way fixed effects specification generates questionable event study estimates that have been only recently brought to light\footnote{See \cite{de_chaisemartin_two-way_2022} for a survey of the literature on problems associated with TWFE models and proposed solutions.} - namely the so called ``negative weights" issue and the possibility of ``cross-lag" contamination laid out in \cite{sun_estimating_2021}. 

In any particular year, the treatment rollout creates three different types of markets, namely the (i) treated-markets, (ii) the not-yet-treated markets, and (iii) the never-treated markets, meaning the markets that never witness an M\&A in the years under observation. If never hosting an M\&A event is correlated to unobserved characteristics of the never-treated markets, then their ability to serve as counterfactuals to treated markets can be compromised. At the same time, if participation in the treated pool is also endogenous to the treated market unobserved characteristics, then the endogeneity related to ever hosting M\&A activity is likely ameliorated by using not-yet-treated markets as controls; after all, the parameter of interest is the average treatment effect on \emph{treated} units, the ATT. For these reasons, I apply the estimation proposed by \cite{callaway_difference--differences_2021} in my investigation. With their method, I can estimate the ATT of merger activity on local labor markets' outcomes using the not-yet-treated units as the counterfactual group and avoid the TWFE specification's  issues reported in the literature. In addition, for the context of mergers, treatment anticipation is a possibility that can be explicitly addressed by their ATT estimation. Next, I expose the identification assumptions from \cite{callaway_difference--differences_2021} in the context of merger activity in local labor markets. 

Let $Y_{m,t}$ denote the outcome variable of local labor market $m$ in year $t$. If an M\&A takes place at market $m$ in a year $g$, the outcome variable is denoted by $Y_{m,t}(g)$, for all years $t$. Begin by considering the parameter

\begin{equation}
    \label{eq:att}
    ATT(g,t) = \mathbb{E}\left[ Y_{m,t}(g) - Y_{m,t}(0) \mid G_g(m) = 1 \right]
\end{equation}
\noindent
where $Y_{m,t}(0)$ is the potential outcome of market $m$ in year $t$ had it not been subject to an M\&A, and $G_g(m)$ is an indicator function that is equal to $1$ if $m$ is treated in year $g$ and $0$ otherwise. $ATT(g,t)$ is what \cite{callaway_difference--differences_2021} call a \emph{group-time average treatment effect}, and in the present case, it represents the expected change in outcome $Y$ in year $t$ for all markets that had an M\&A event in year $g$, e.g., $ATT(2008,2012)$ is the $ATT$ value in 2012 for markets treated in 2008. The group-time average treatment effects can be combined into coefficients that recover event-study type estimates \citep{callaway_difference--differences_2021}. More precisely, the expected change in outcome $Y$ at market $m$ after $l$ periods of exposure to treatment is given by 

\begin{equation}
    \label{es_coeff}
    \beta^{Y}_{ES}(l) = \sum_{g\in \mathcal{G}}\mathbbm{1}\left\{g+l\leq\mathcal{T}\right\}P(G=g \mid g+l\leq\mathcal{T})ATT(g,g+l)
\end{equation}

\noindent
where $\mathcal{G}$ is the set of all treatment years and $\mathcal{T}$ is the last year of observation. The term $P(G=g \mid g+l\leq\mathcal{T})$ weighs all the group-time $ATT(g,t)$ in which $t$ is observed, and $t=g+l$. The more units in a treatment cohort, the more weight that cohort gets in the event study coefficient $\beta^{Y}_{ES}(l)$.   

\subsubsection{Identification}
The causal identification of $ATT(g,t)$ in Equation (\ref{eq:att}) relies on a modified version of the parallel trends' assumption used in canonical $2 \times 2$ difference-in-differences settings. This and all the other assumptions needed for identification are presented below. 

\begin{description}
    \item[Assumption 1](Irreversible Treatment) For $t=1,\dots,\mathcal{T}$, $G_1(m) = 0$ almost surely (a.s.), and, if $G_t(m)=1$, then $G_{t'}(m)=1$ for all $t'>t$ almost surely.    
    
    \item[Assumption 2](Random Sampling) Treatment status and outcome variables of individual markets are independent and identically distributed, i.e. $\left\{Y_{m,t},G_t(m)\right\}_{(m,t)=(1,1)}^{(M,\mathcal{T})}$ is $iid$.  
    
    \item[Assumption 3 \label{ass:anticipation}](Treatment Anticipation of $\zeta$ Years) Let $\mathcal{G}$ be the set of all years of treatment. For all $t<g-\zeta$ and $g\in \mathcal{G}$, 
    \begin{equation}
        \mathbb{E}\left[Y_{m,t}(g) \mid G_g(m) = 1 \right] = \mathbb{E}\left[Y_{m,t}(0) \mid G_g(m) = 1 \right] \text{ a.s.}
    \end{equation}
    
    \item[Assumption 4](Parallel Trends Based on ``Not-Yet-Treated" Units) For all $g \in \mathcal{G}$, $t \geq g-\zeta$, and $s>t+\zeta$, the following equality holds a.s. 
    \begin{equation}
    \label{parallel_trends}
        \mathbb{E}\left[Y_{m,t}(0) - Y_{m,t-1}(0)\mid G_g(m)=1 \right] = \mathbb{E}\left[Y_{m,t}(0) - Y_{m,t-1}(0)\mid G_g(m)=0\text{, }G_s(m)=0\right]
    \end{equation}
    \item[Assumption 5](Overlap) For each $t \in \{2,\dots,\mathcal{T}\}$, and group $g$ in $\mathcal{G}$, at least one market is treated in year $g$, and for years $g+t$, at least one market remains untreated.  
\end{description}

\begin{figure}[htp]
    \centering
    \caption{Years With Merger Activity per Market and Treatment Status}
    \begin{subfigure}[C]{.7\textwidth}
        \centering
        \caption{Years With Merger Activity per Market}
        \includegraphics[width=\textwidth]{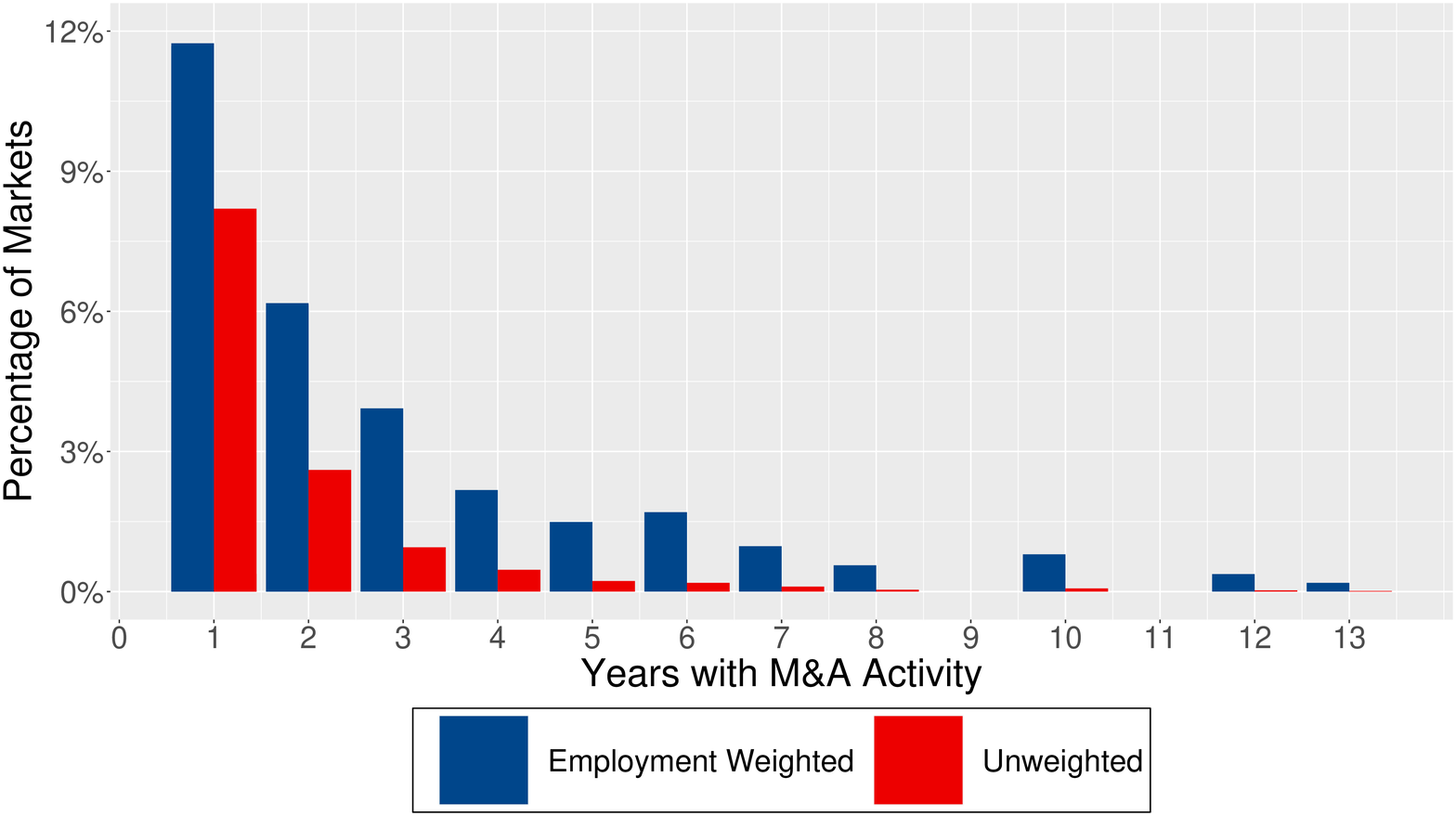}
        \label{fig:g_hits_trade_secc}
    \end{subfigure}
    \vfill
    \begin{subfigure}[C]{.7\textwidth}
        \centering
        \caption{Treatment Status}
        \includegraphics[width=\textwidth]{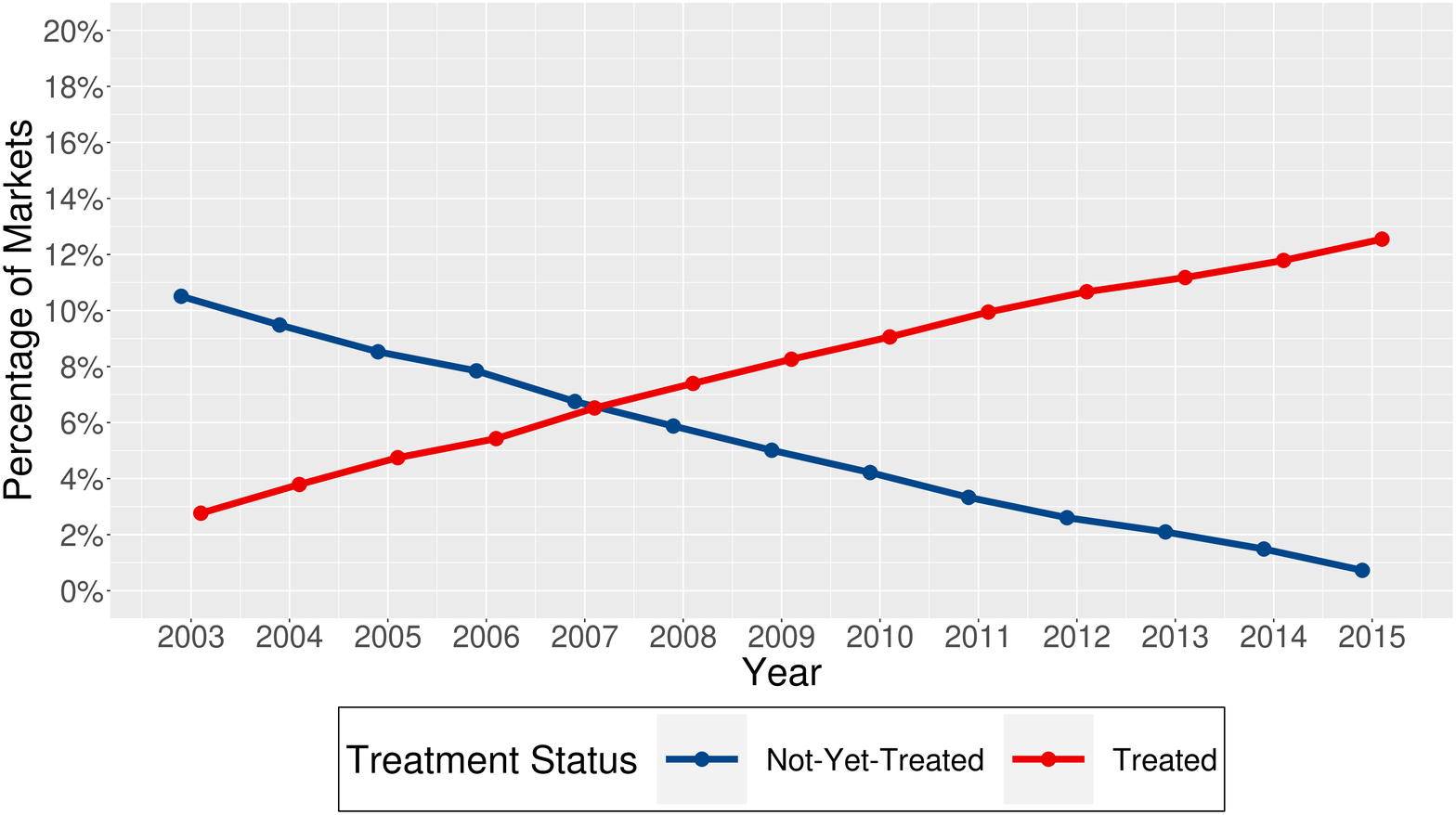}
        \label{fig:g_status_first_trade_secc}
    \end{subfigure}
    \label{fig:llm_status}
    \vspace{-0pt}
    \begin{minipage}{.7\textwidth}
    \small  
    Note: Local labor markets are defined by pairs of commuting zone and 3-digit industry code. The Not-Yet-Treated status refers to markets that eventually witness an M\&A event in the following observed years. Most markets never experience an M\&A event. The second most common case is that of markets with only one year of merger activity.             
    \end{minipage}
\end{figure}

Assumption 1 imposes that once a local labor market is treated with an M\&A, it remains treated throughout the analysis and does not contribute again to the control group of the estimation. For markets with multiple M\&A events in different years, I will use the first event as the treatment date. Most markets never have an M\&A in the years of observation. Following that, out of any particular number of years with merger activity, the second most common case is that of markets with only one year of M\&A events (Figure \ref{fig:g_hits_trade_secc}) \footnote{Markets with one or multiple M\&As in the same year are equally considered to be treated in that year.}.  Assumption 2 is met by the use of a balanced panel. Assumption 3 is a relaxation of the canonical \emph{no treatment anticipation} condition in $2\times 2$ DiD setups. Here, one can allow the treatment to begin as far back as the context requires, effectively resulting in two changes: (i) moving the more commonly used normalization point in event studies from $t=-1$ to $t=-\zeta-1$; and (ii), removing the units already under anticipated treatment dynamics from the control units pool. Mergers and Acquisitions can be lengthy processes, and the news of an M\&A might induce changes in the behavior of firms and workers before the reported year of the event. I present results with no anticipation of treatment, i.e. $\zeta=0$, but I also find robustness to $\zeta=1$. 

Assumption 4 states the main requirement for identification, namely that the year-over-year change in the outcome of treated markets, had they not been treated with an M\&A, is the same as that of markets that will eventually get treated in the following years—the \emph{not-yeat-treated} units. In conjunction with Assumption 3, the parallel trends assumption imposes that, in year $t$, the counterfactual of previously treated markets are the ones that get treated in years later than $t+1+\zeta$ – e.g., for a given treated market $m$, and anticipation $\zeta=0$, its counterfactual in year, let's say, 2010, corresponds to all markets that will be treated in 2011 or later; notice that, in this example, markets treated in 2010 are already subject to treatment dynamics, and therefore cannot be used as controls to treated markets in 2010.   

Figure \ref{fig:g_status_first_trade_secc} shows the percentage of markets by treatment status every year according to the occurrence of the first M\&A event. The majority of labor markets have no M\&A activity throughout the years of observation. For pairs of commuting zone and 3-digit tradable industry codes, the share of never-treated markets is 86.72\% - or 7,526 out of 8,678 markets. The remaining markets are either treated or not-yet-treated, depending on the year. In addition, Figure \ref{fig:g_status_first_trade_secc} shows that in all years, there is a positive fraction of markets that remains untreated, on top of never there being a year with 100\% of markets treated, which contemplates Assumption 5. At the same time, the trajectory of the fraction of Treated and Not-Yet-Treated markets is smooth across the years, not indicating any noticeable discontinuities that could demand inference by different time windows. As is common in event studies, fewer markets remain untreated towards the end of the timeline under observation. For this reason, I'll restrict the analysis to a window of 5 years around the merger event. Standard assumptions guarantee the estimation of cluster-robust standard errors by means of a multiplier bootstrap procedure   \footnote{An exposition of the assumptions necessary for inference is beyond the scope of this paper, and I refer the reader to Section 4, page 211, in \cite{callaway_difference--differences_2021}.}.    

\subsubsection{The Estimand and Control Groups}
Under the assumptions outlined above, one can rewrite the treatment effect in \ref{eq:att} as
\begin{equation}\label{eq:estimand}
    ATT(g,t) = \mathbb{E}\left[Y_{m,t} - Y_{m,g-1}\mid G_g(m)=1 \right] - \mathbb{E}\left[Y_{m,t} - Y_{m,g-1}\mid G_{t+1}(m)=0 \right]
\end{equation}

\noindent
effectively obtaining the main estimand used in this paper. Let me take a moment to describe the terms of the expression in Eq. \ref{eq:estimand}. Suppose that a group of local labor markets had M\&As in year $g$. The average treatment effect on this group, at any year $t$, the value $ATT(g,t)$, is equivalent to the difference in the average increment of their outcome variable since one year before treatment, i.e., year $g-1$, and the average increment of all markets not yet subject to an M\&A by year $t+1$, relative to that same base year $g-1$. In other words, the 2012 wage effect on a labor market treated in 2008, noted by $ATT(2008,2012)$, is identified by the difference between the wage growth since 2007 of all markets treated in 2008, and the wage growth among eventually treated markets that remain not treated by 2013. Notice that, in the case of allowing for a positive time length of treatment anticipation, $\zeta=1$, for instance, the markets used in the control group would have to be those that remain untreated up to 2014, i.e., two years later than the year at which the effect is being measured, also moving the reference year from 2007 to 2006.     

%--------------------------------------------------------

\subsection{The Earnings Variable}

One of the possible consequences of mergers is the change in workforce composition within merged firms, and ultimately in the whole labor market. This is specially relevant if, for instance, bigger firms are able to hire younger less skilled workers to replace more experienced, more educated, and thus more costly ones. Any estimated decline in wages would then be the result of turnover towards employees with less attractive observable attributes, and not necessarily due to market-wide changes in the competition for labor services. Given the details about workers' observed attributes in the data, I will estimate the effects of mergers on a measure of earnings that takes such attributes into account, thus obtaining an effect that is not driven by changes in the composition of attributes of workers. Similar to \cite{arnold_mergers_2022}, I estimate local labor market-level wages that control for workers characteristics available in the data. The parameter I look for is $\theta_{m,t}$ in

\begin{equation}
    \label{eq:llm_wage}
    w_{i,m,t} = \theta_{m,t} + \beta_t X_{i,t} + u_{i,m,t}
\end{equation}

\noindent
where $w_{i,m,t}$ is worker $i$'s log annual earnings in market $m$ and year $t$, and $X_{i,t}$ is their vector of observable attributes \footnote{$X_{i,t}$ contains dummies for race, college and high school diplomas, sex, and a quadratic binomial on age.}. This model is estimated via OLS for every year in the data. This way, $\theta_{m,t}$ captures the annual market-level log wage net of trends in the workforce composition $X_{i,t}$\footnote{Later on, in Section \ref{sec:results}, when I split the analysis between spillover and merged firms, their respective earnings measures are obtained from a re-estimation of Equation \ref{eq:llm_wage} for each group separately.}.

%-------------------------------------------------------------
\subsection{Summary Statistics}

Table \ref{tab:all_mkts} presents summary statistics from the pool of eventually treated local labor markets, where all means are computed based on the year prior to the first M\&A event in the market. Despite the apparent large number of firms per local labor market (almost 63 on average), the HHI score shows that employment is unevenly distributed among employers – 2,847.50 is above the threshold of 2,500 used in DOF-FTC Horizontal Merger Guidelines to consider a product market highly concentrated. This score is equivalent to a market with 3.51 equally sized employers, and is above the overall measured concentration in the U.S. (around 1,500 \citep{rinz_labor_2020}), but below the one found looking only at manufacturing (3,380 \citep{benmelech_strong_2022}).         

\begin{table}[]
\centering
\caption{\label{tab:all_mkts}Summary Statistics on Local Labor Markets}
\vspace{-12pt}
\adjustbox{max width=\textwidth}{%
\begin{threeparttable}
\begin{tabular}{lrlrllll}
 & \multicolumn{1}{l}{}          &  & \multicolumn{1}{l}{}     &  &                               &  &  \\ \hline
 & \multicolumn{1}{l}{}          &  & \multicolumn{1}{c}{Mean} &  & \multicolumn{1}{c}{Std. Dev.}  &  &  \\ \hline
 & \multicolumn{1}{l}{}          &  & \multicolumn{1}{l}{}     &  &                               &  &  \\
 & Annual Earnings \small{(in 2010 BRL)}         &  & 3,905.40                &  & \multicolumn{1}{r}{1,717.94} &  &  \\
 & Workers                       &  & 1,727.10                 &  & \multicolumn{1}{r}{2,618.95}  &  &  \\
 & Workers in Merging Firms (\%) &  & 39.66                    &  & \multicolumn{1}{r}{32.89}     &  &  \\
 & Firms                         &  & 62.91                    &  & \multicolumn{1}{r}{142.59}    &  &  \\
 & HHI                           &  & 2,847.50                 &  & \multicolumn{1}{r}{2,549.25}  &  &  \\
 & \multicolumn{1}{l}{}          &  & \multicolumn{1}{l}{}     &  &                               &  &  \\
 & Market-Year Observations      &  & 10,348                   &  &                               &  &  \\ \hline
 & \multicolumn{1}{l}{}          &  & \multicolumn{1}{l}{}     &  &                               &  & 
\end{tabular}
\begin{tablenotes}
   \vspace*{-11pt}
   \footnotesize
   \item\textbf{Notes:} These are summary statistics of local labor market characteristics. Local labor markets are defined by pairs of commuting zone and 3-digit industry sector codes. Annual earnings were adjusted by inflation, with base year in 2010. All means are computed one year before the M\&A event.      
   \end{tablenotes}
   \end{threeparttable}}
\end{table}

%---------------------------------------------------------------------------------------------------

\section{\label{sec:results}Results}

In this section, I present the study events as well as overall estimates of the treatment effects of M\&As on workers earnings, log employment, and concentration measured by HHI. I also report estimates from among only the firms that constitute the merger deal, and all other firms in the same labor market. Finally, I explore the role of employment concentration on the treatment effects by comparing the estimates from events with predicted zero impact on HHI - the out-of-market mergers, and those on the top of the distribution of concentration increases – the high-impact mergers.  

%--------------------------------------------------------

\subsection{\label{sec:res_all}Market-level Effects}

\begin{figure}[htp]
\begin{adjustwidth}{-1.5cm}{-1.5cm}
    \centering
    \caption{ATT of Earnings, Employment, and HHI}
    \begin{subfigure}[t]{0.49\linewidth}
        \centering
        \caption{Workers Earnings}
        \includegraphics[width=\textwidth]{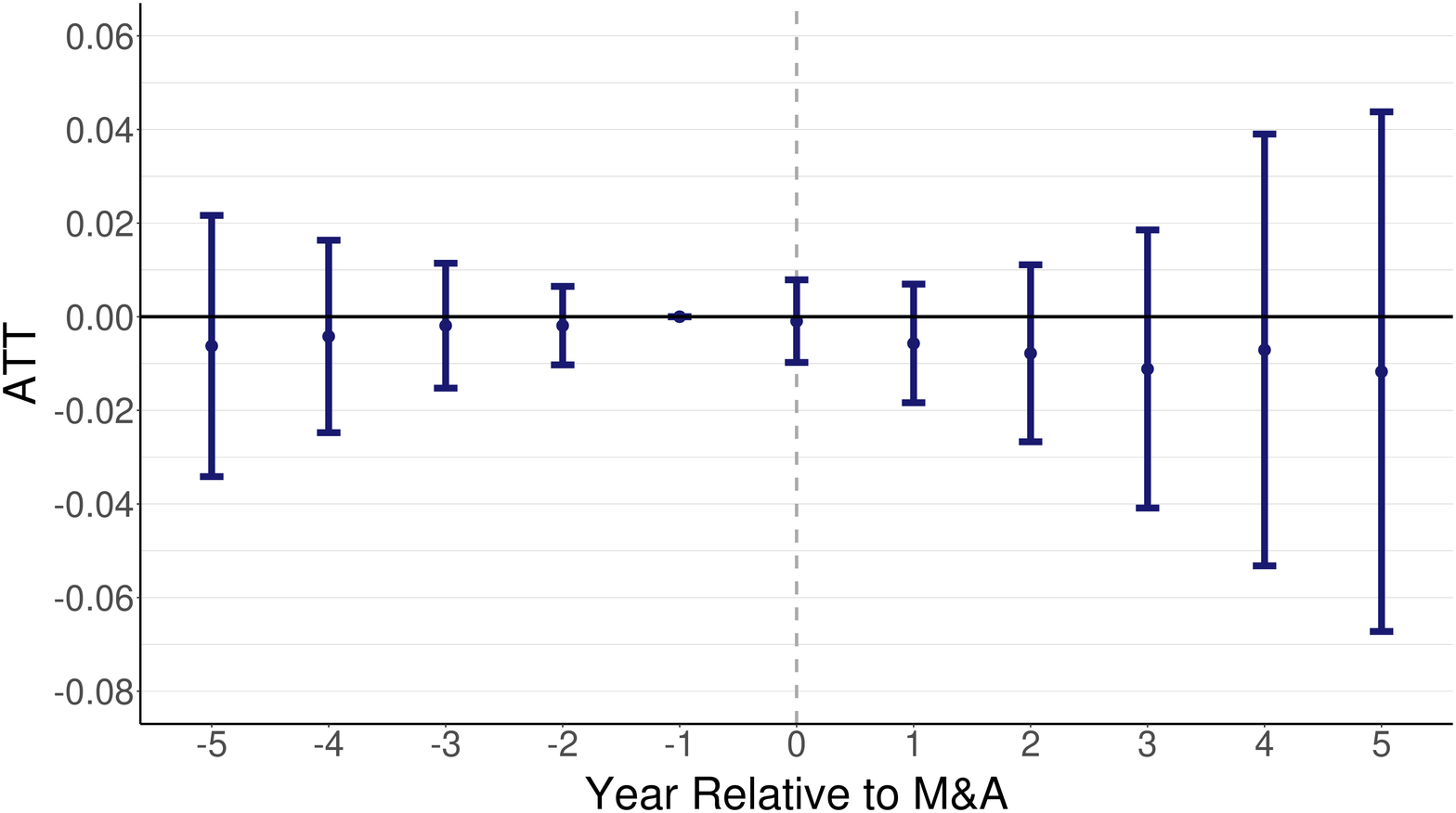}\label{fig:g_mod_w_nyt_llm_secc3_fe_tradable_first_tradable_0}
    \end{subfigure}
    \hfill
    \begin{subfigure}[t]{0.49\linewidth}
        \centering
        \caption{Log Employment}
        \includegraphics[width=\textwidth]{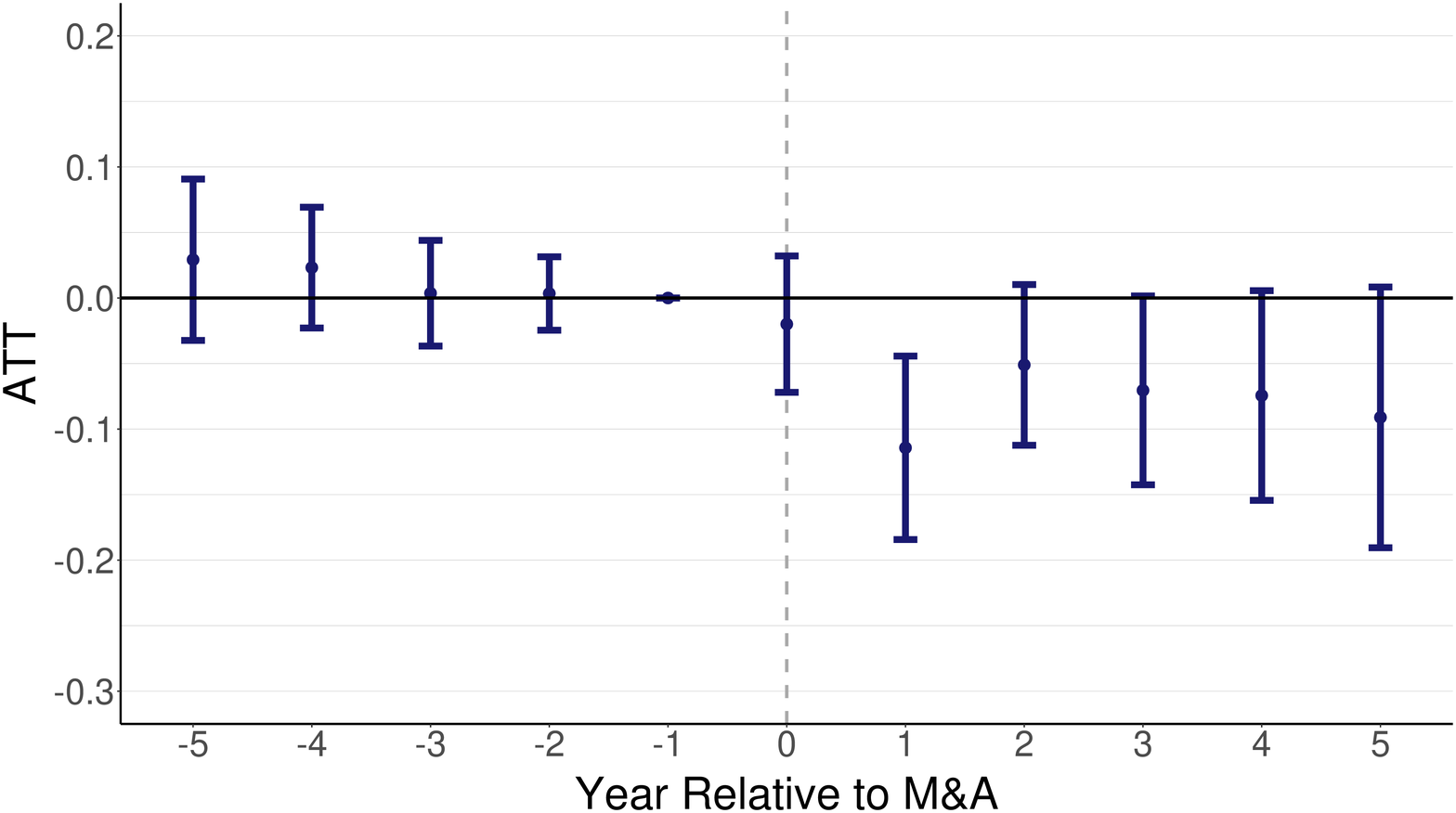}\label{fig:g_mod_w_nyt_llm_secc3_ur_lqt_all_first_tradable_0}
    \end{subfigure}
    
    \vspace{0.6cm}
    
    \begin{subfigure}[t]{0.49\linewidth}
        \centering
        \caption{Employment HHI}
        \includegraphics[width=\textwidth]{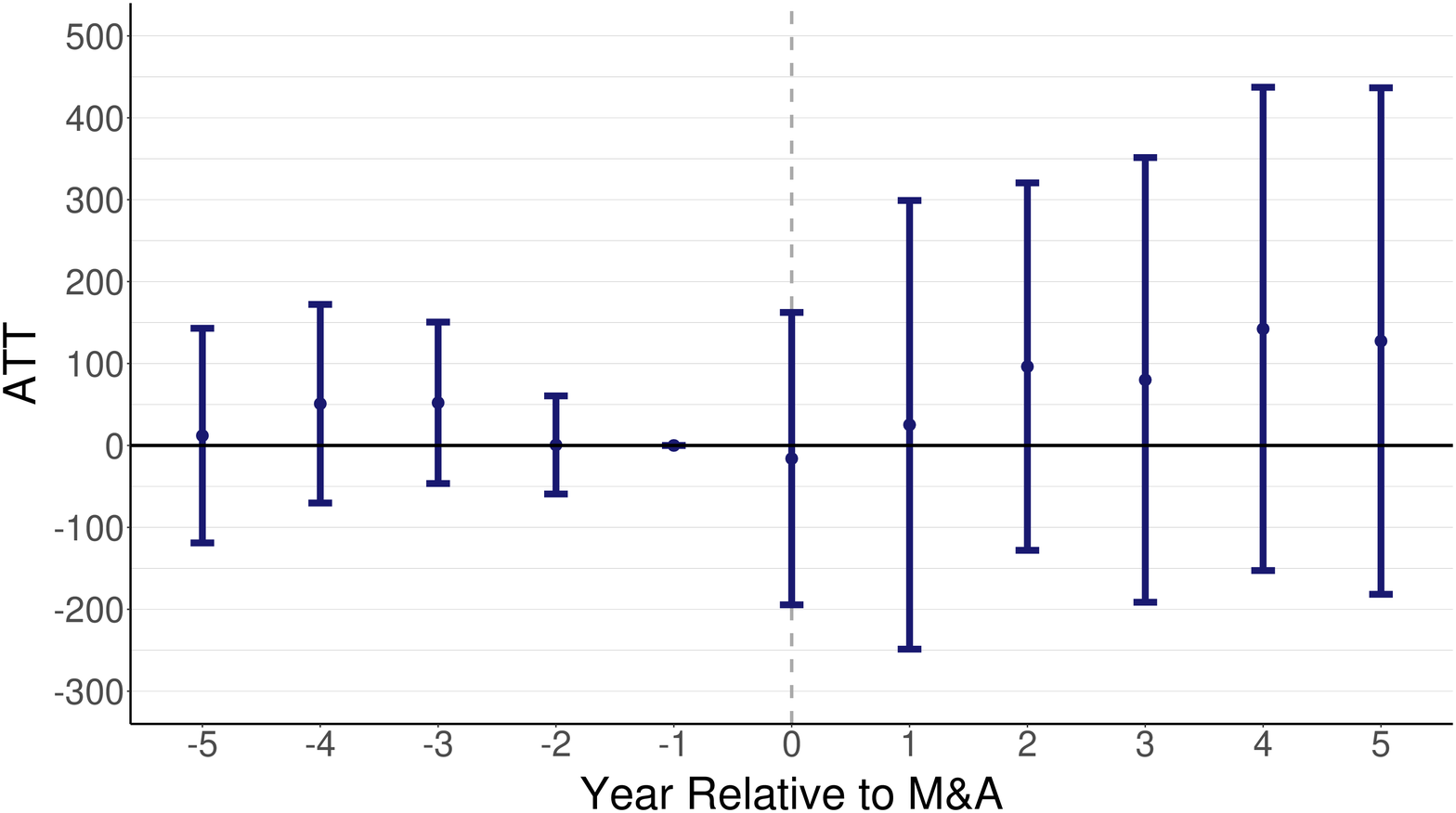}\label{fig:g_mod_w_nyt_llm_secc3_ohhi_first_tradable_0}
    \end{subfigure}
    \label{fig:g_first_all3}
\end{adjustwidth}
\vspace{.25cm}
 
    \scriptsize 
    \textit{Note:} 95\% confidence intervals with standard errors clustered at the market level. The year $t=0$ marks the first  M\&A in a local labor market, defined by pairs of commuting zone and 3-digit level industry code. Markets are weighted by their relative employment size in each year. 

\end{figure}

Figure \ref{fig:g_first_all3} shows the event study estimates on earnings, employment, and concentration measured by HHI. In years before $t=0$, the date of the M\&A event, the outcome variables trend similarly among treated and not-yet-treated labor markets, which supports the plausibility of the parallel trends assumption. In the years post-M\&A, I cannot rule out that market-level earnings of workers in treated local labor markets have changed on par with the earnings of workers in not-yet-treated units, despite the negative point estimates. In the next section, I estimate separate effects for the treated firms, i.e., those firms that took part in the M\&A, and all other firms in the same market. This distinction will shed light on the source of negative earnings estimates and possible market interactions responsible for generating these results. 

Notwithstanding the stability in earnings differences, the effect of M\&As is negative on employment, measured in log, corresponding to $-0.1143$ ($SE=0.0249$) one year after the event. The remaining lag estimates indicate that the level of employment does not recoup to its counterfactual trajectory. Overall, in the five years post-treatment, there is a 6.78 percent relative decline in employment on treated markets\footnote{This percentage effect is obtained from exponentiating the overall measured effect, $1-\exp(-0.0702)$.}. The fall in employment takes place even when the estimates show a close to null change in employment concentration, as can be seen in Panel \ref{fig:g_mod_w_nyt_llm_secc3_ohhi_first_tradable_0}. M\&As are expected to increase local concentration mechanically, but the dynamic estimates show that the workers in treated markets are not subject to a significantly more concentrated labor market after the M\&A.          

%-----------------------------------------------------------------------------------------------------

\subsection{\label{sec:res_split}Effects from Within Merged Firms and Spillover Dynamics}

\emph{A priori}, there is no reason to expect that all employers in a labor market will be equally impacted by a merger event, or that employment and wage adjustments will be similar across all workers. The merging firms might experience changes intrinsical to the merger that might not propagate to the rest of the labor market as a whole. I will now, on one hand, compare the earnings and employment effects only among firms that participate in the merger. On the other, I look at the earnings and employment from all other firms within the same labor market, which I will call spillover firms from now on. The observation of the two separate outcomes, one for within-merged and another for spillover firms, sheds light on what can be the market-wide effect of the M\&As and what is related to unobserved changes mergers enact in their entities. It is possible to expect that the new ownership leads to changes in managerial practices, or even in worker productivity, both of which can affect earnings and employment within the newly merged firm, but not necessarily those of competitors in the same labor market.  

\begin{figure}[htp]
\begin{adjustwidth}{-1.5cm}{-1.5cm}
    \centering
    \caption{Effects from Within and Spillover Firms}
    \begin{subfigure}[t]{0.49\linewidth}
        \centering
        \caption{Workers Earnings}
        \footnotesize \emph{Within M\&A Firms}
        \includegraphics[width=\textwidth]{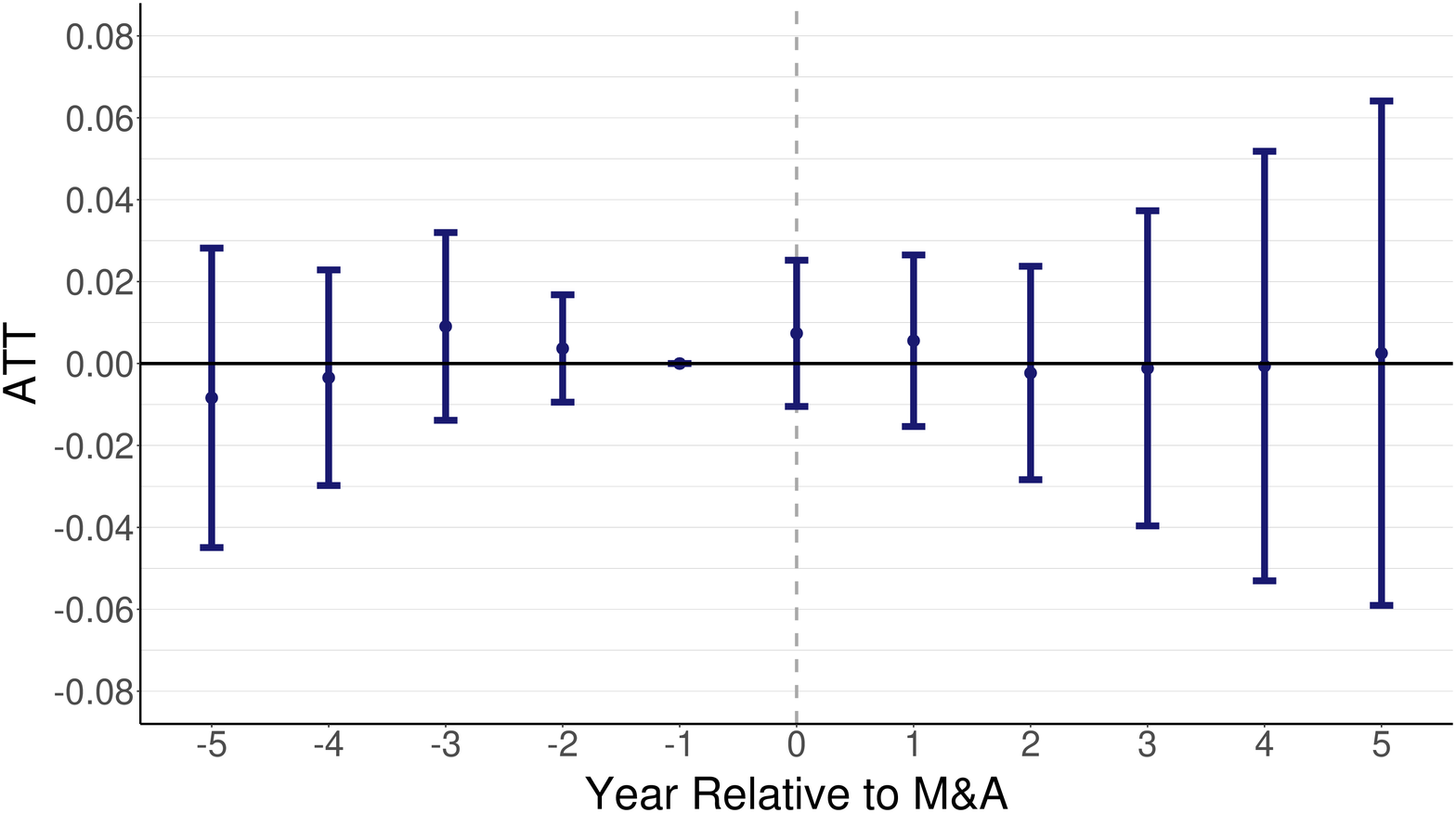}
        \label{fig:g_mod_w_nyt_llm_secc3_fe_spill0_tradable_first_tradable_0}
    \end{subfigure}
    \hfill
    \begin{subfigure}[t]{0.49\linewidth}
        \centering
        \caption{Workers Earnings}
        \footnotesize \emph{Spillover Firms}
        \includegraphics[width=\textwidth]{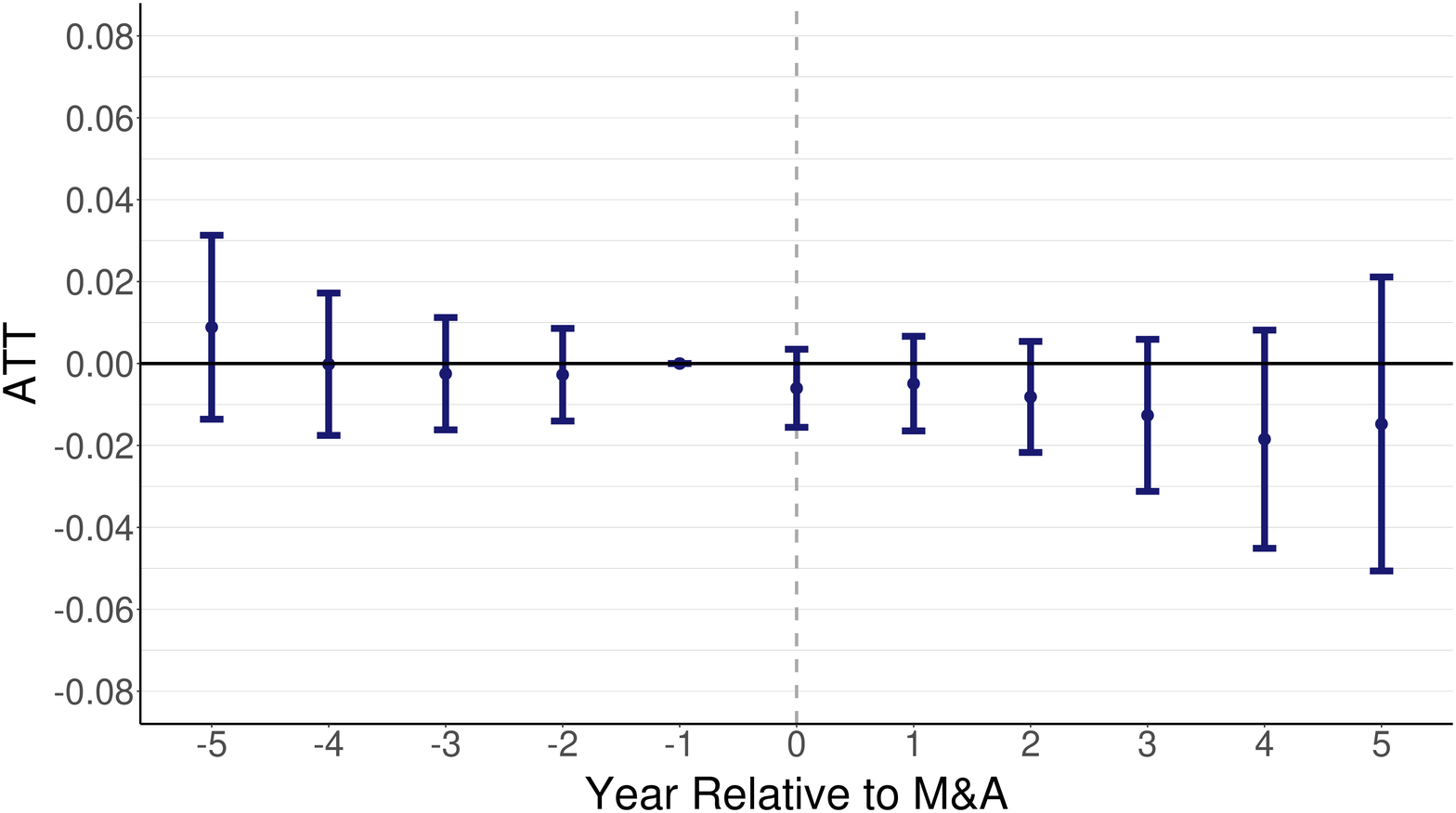}
        \label{fig:g_mod_w_nyt_llm_secc3_fe_spill1_tradable_first_tradable_0}
    \end{subfigure}
    \vspace{0.2cm}
    \begin{subfigure}[t]{0.49\linewidth}
        \centering
        \caption{Log Employment}
        \footnotesize \emph{Within M\&A Firms}
        \includegraphics[width=\textwidth]{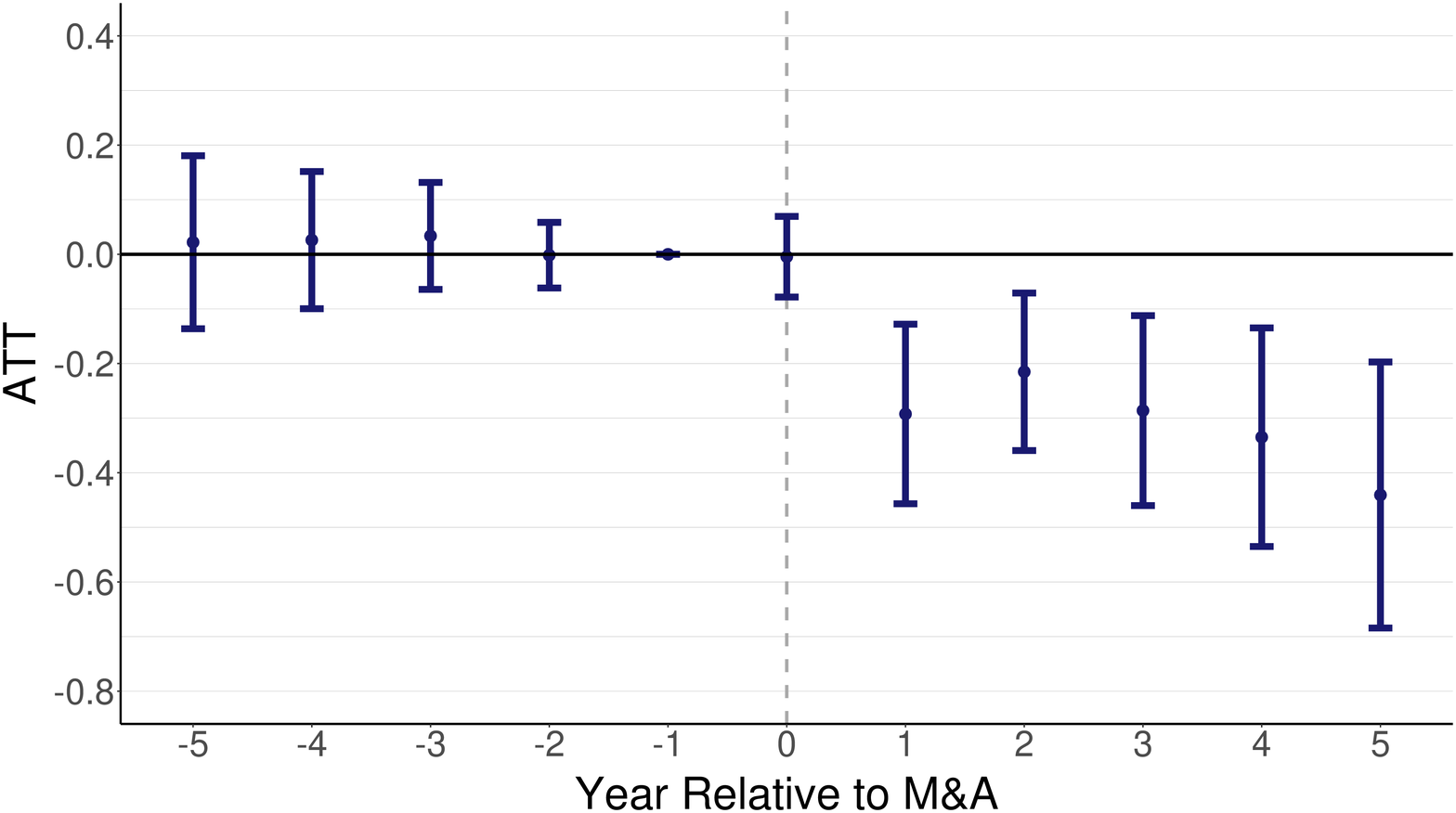}
        \label{fig:g_mod_w_nyt_llm_secc3_ur_lqt_spill0_first_tradable_0}
    \end{subfigure}
    \hfill
    \begin{subfigure}[t]{0.49\linewidth}
        \centering
        \caption{Employment}
        \footnotesize \emph{Spillover Firms}
        \includegraphics[width=\textwidth]{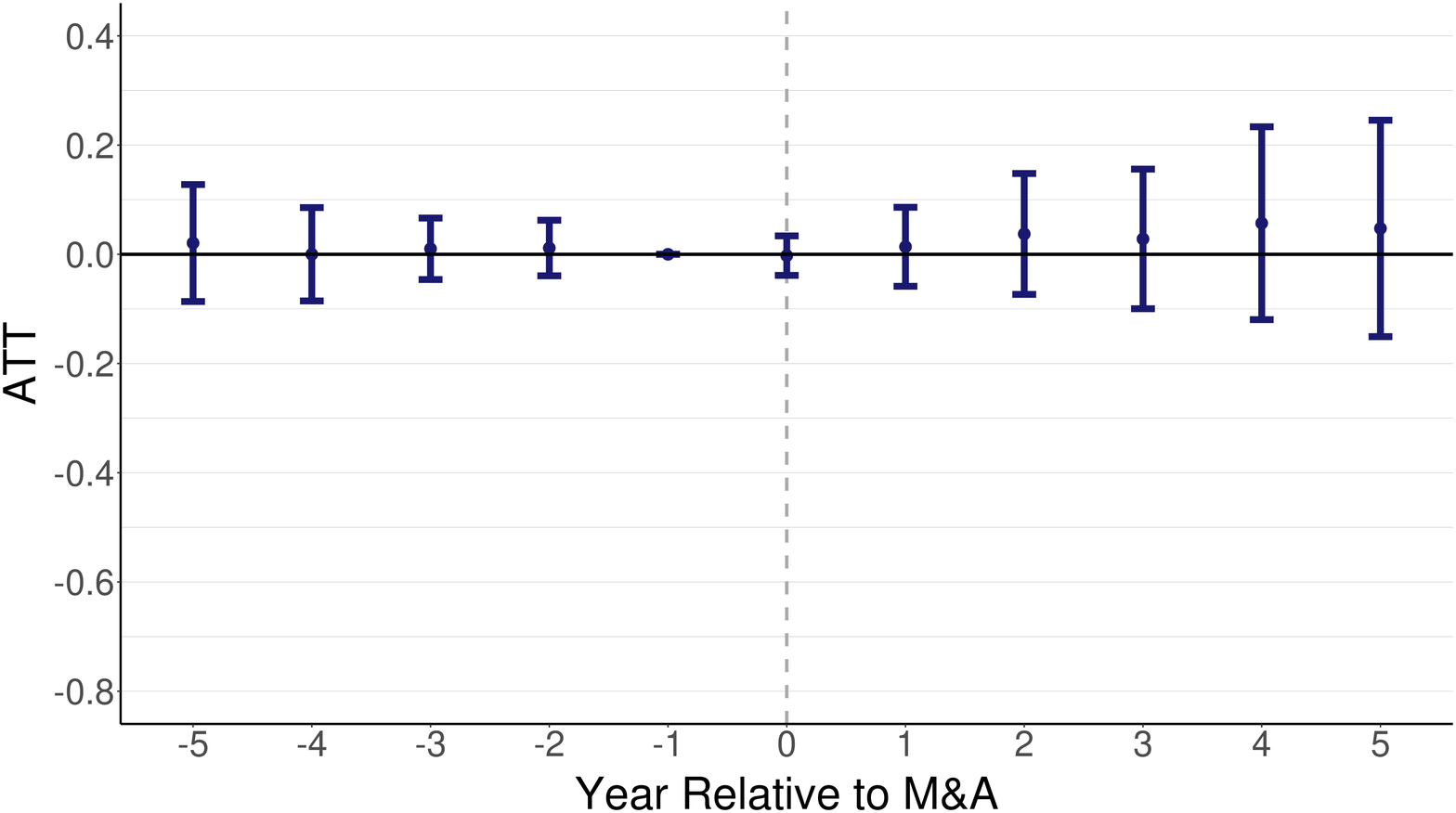}\label{fig:g_mod_w_nyt_llm_secc3_ur_lqt_spill1_first_tradable_0}
    \end{subfigure}
    \vspace{0.2cm}
    \begin{subfigure}[t]{0.49\linewidth}
        \centering
        \caption{Log Hires and Log Separations}
        \footnotesize \emph{Within M\&A Firms}
        \includegraphics[width=\textwidth]{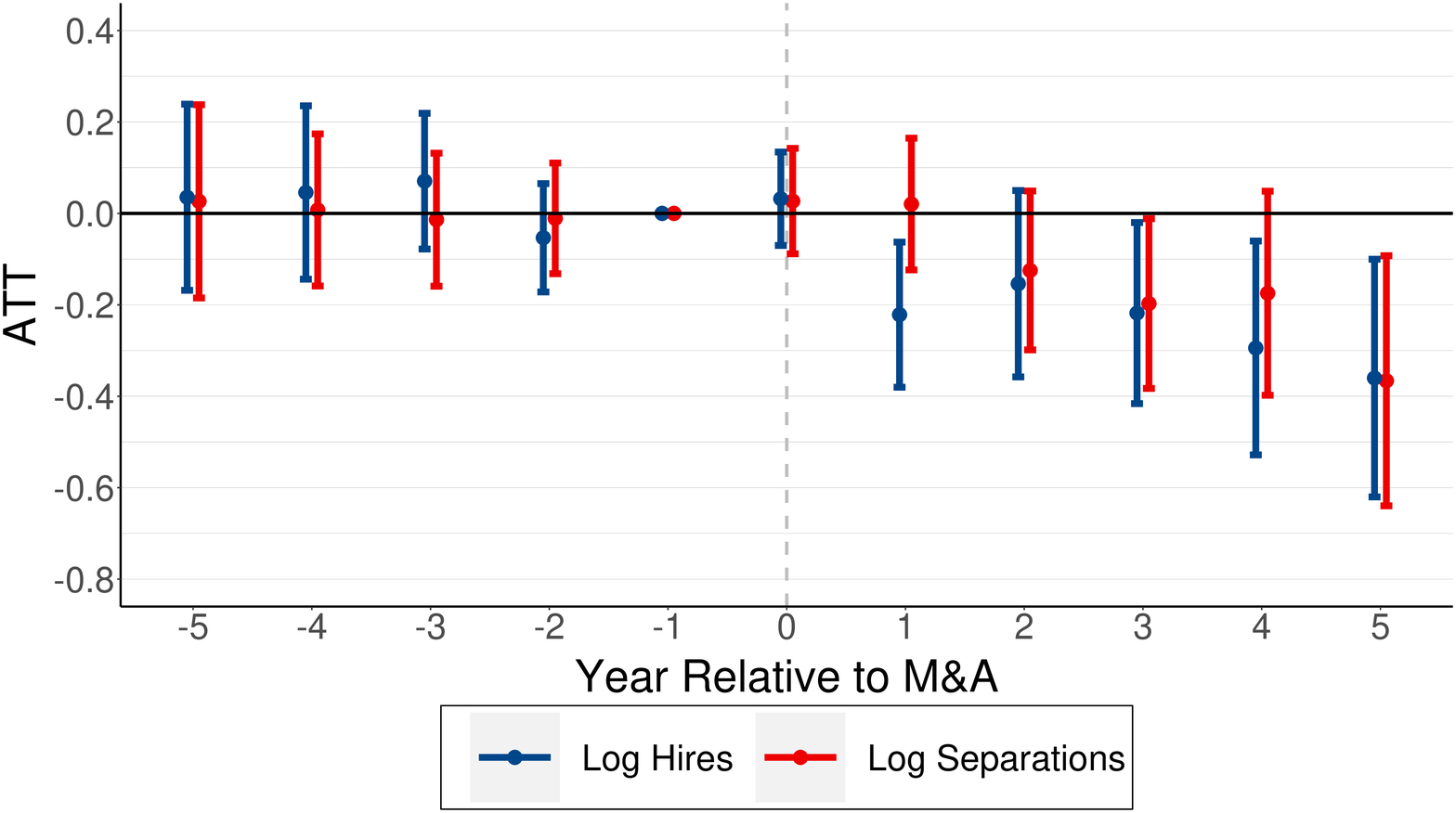}
        \label{fig:g_mod_w_nyt_llm_secc3_lseparations_spill0_first_tradable_0}
    \end{subfigure}
    \hfill
    \begin{subfigure}[t]{0.49\linewidth}
        \centering
        \caption{Log Hires and Log Separations}
        \footnotesize \emph{Spillover}
        \includegraphics[width=\textwidth]{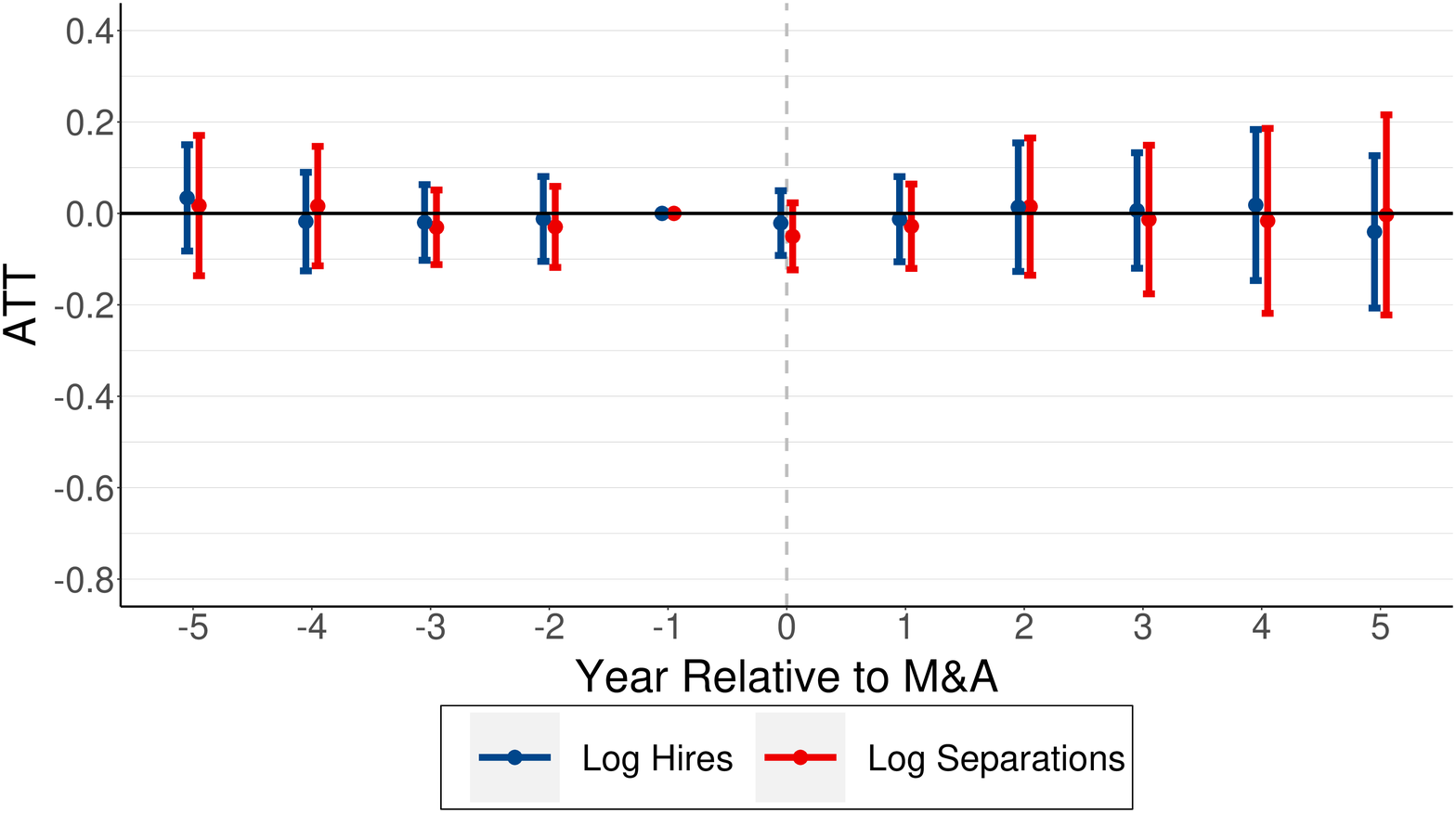}
        \label{fig:g_mod_w_nyt_llm_secc3_lseparations_spill1_first_tradable_0}
    \end{subfigure}
    \label{fig:g_within_spill}
\end{adjustwidth}
\vspace{-0cm}

    \scriptsize 
    \textit{Note:} 95\% confidence intervals with standard errors clustered at the market level. The year $t=0$ marks the first tradable M\&A in a local labor market, defined by pairs of commuting zone and 3-digit level industry code. Tradable M\&As are mergers and acquisitions of tradable industry sectors. Spillover effects are obtained at the local labor market level by excluding the workers employed in the merged or acquired firms. Incumbents are employees with at least 30 months of tenure in their current job by the end of each year. New hires are workers with less than 12 months of tenure by Dec. 31 of each year. Markets are weighted by their relative size in each year.   

\end{figure}

\begin{figure}[htp]
\begin{adjustwidth}{-0cm}{-0cm}
    \centering
    \caption{\label{fig:g_ov_all}Overall Effects}
    \begin{subfigure}[t]{0.45\linewidth}
        \centering
        \caption{Workers Earnings}
        \includegraphics[width=\textwidth]{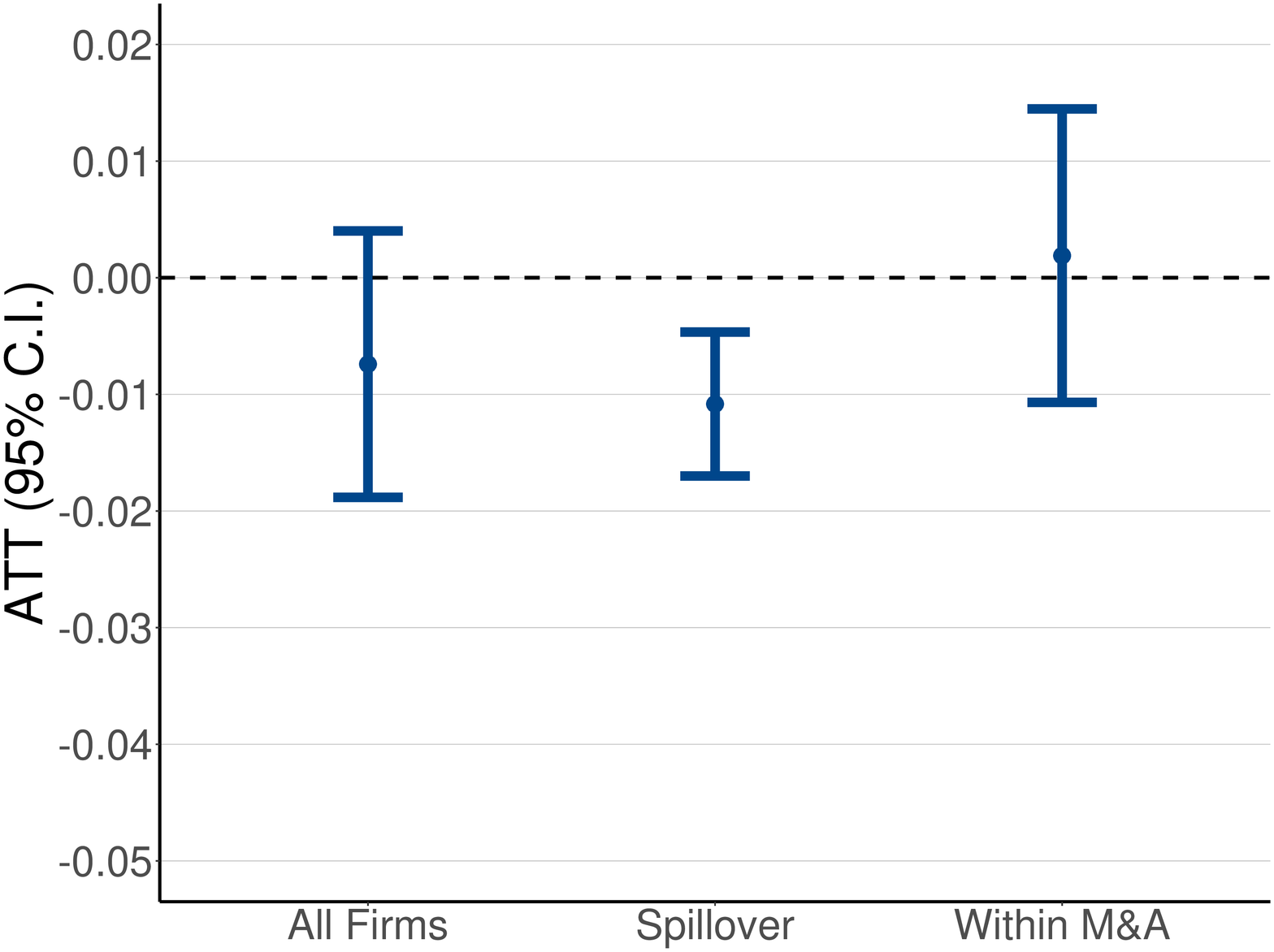}
        \label{fig:g_ov_earn_all}
    \end{subfigure}
    \hfill
    \begin{subfigure}[t]{0.45\linewidth}
        \centering
        \caption{Log Employment}
        \includegraphics[width=\textwidth]{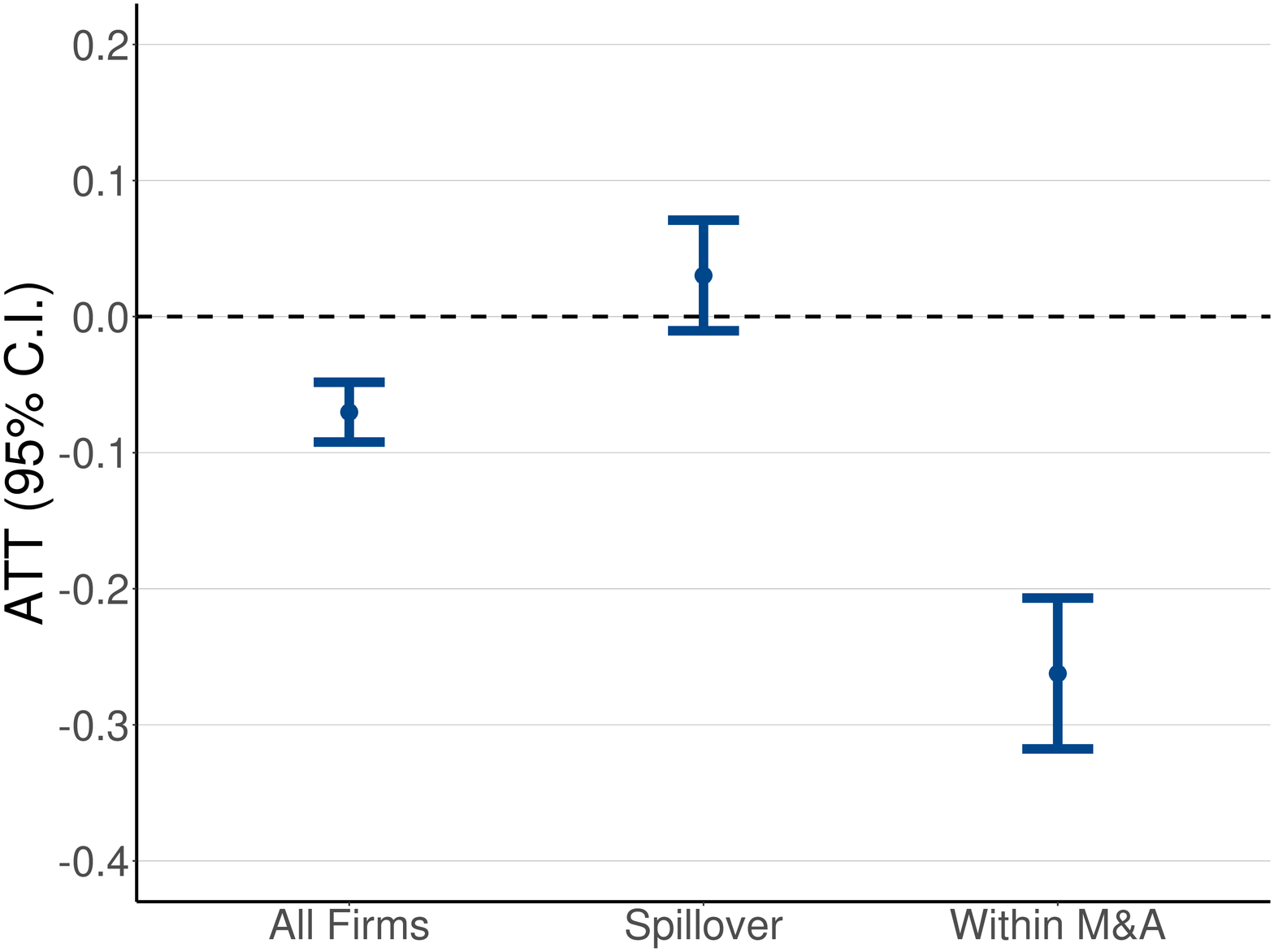}\label{fig:g_ov_emp}
    \end{subfigure}
    \label{fig:g_ov_emp_all}    
\end{adjustwidth}
\vspace{-0cm}
    
    \scriptsize 
    \textit{Note:} The graph shows the ATT for all five years after treatment. 95\% confidence intervals with standard errors clustered at the market level. Markets are defined by pairs of commuting zone and 3-digit level industry code. Markets are weighted by their relative size each year. 
    
\end{figure}

Figure \ref{fig:g_within_spill} shows the event study estimates of earnings and employment effects of the merged and spillover firms separately. Panel \ref{fig:g_mod_w_nyt_llm_secc3_fe_spill0_tradable_first_tradable_0} shows that earnings in merged firms do not diverge from the earnings in not-yet-merged firms in other labor markets, with lag point estimates close to zero, especially after the second year of treatment exposure. Differently, earnings in spillover firms decline after the merger. The 95 percent confidence intervals contain zero for the separate lag estimates, but the overall effect five years after treatment represents a 1.07 percent decline in wages (Figure \ref{fig:g_ov_all}). The employment estimates show a different dynamic. Compared to their baseline difference from other merger participants in not-yet-treated markets, employment in merged firms is significantly lower. The estimate of one year of exposure to the M\&A is -0.2923 (SE=0.0052), and, for all five years after the event, I estimate a 23.07 percent decrease in employment in merged firms (Figure \ref{fig:g_ov_all}). At the same time, I fail to reject zero employment effects in spillover firms – although the overall effect is positive but imprecisely estimated (Figure \ref{fig:g_ov_all}). Taken together, these findings show that the negative employment effects shown in Figure \ref{fig:g_first_all3} were carried out primarily by the firms participating in the merger, while the negative wage point estimates originated from a decline in earnings in spillover firms, with a positive but, not significant, increase in employment of spillover firms. The bottom two panels in Figure \ref{fig:g_within_spill} show the estimates for hiring and separations of each type of firm. Panel \ref{fig:g_mod_w_nyt_llm_secc3_lseparations_spill0_first_tradable_0} shows that the negative employment in merged firms is adjusted via an abrupt decline in hires, while separations only start declining after the first year of exposure to treatment. Panel \ref{fig:g_mod_w_nyt_llm_secc3_lseparations_spill1_first_tradable_0} reinforces the findings for employment in spillover firms, showing that both hires and separations remain similar to their \emph{pre}-merger levels.       

%--------------------------------------------------------

\subsection{\label{sec:res_hi_low} Effects from M\&As with Different Concentration Changes} 
There are many channels through which mergers can induce changes in wages and employment levels in local labor markets. On the side of merged firms themselves, one can think that new management or changes to worker productivity might be the cause of the observed outcomes. At a market level, on the other hand, changes in the competition for labor services might be the driver of the decline in wages and employment, especially if mergers foster an anticompetitive behavior from employers. Recent studies that look at wage and employment effects of mergers have found that labor market concentration is an important mediator of the relationship between employer consolidation and earnings declines. Using U.S. hospital mergers, \cite{prager_employer_2021} find that only consolidations in the top quartile of local concentration shocks induce negative wage effects on health sector workers. In a more general setting, using a similar definition of local labor market as the one in this paper, \cite{arnold_mergers_2022} finds that mergers below the 80th percentile of predicted concentration change do not have significant impact on workers' earnings. The fact that larger negative wage effects are found only when mergers enact larger shifts in concentration is consistent with the oligopsony theory à la Cournot of competition in labor markets \citep{azar_measuring_2019, arnold_mergers_2022}. In such models, the first-order condition of each firm's profit maximization problem can be combined to obtain a negative relationship between market wages and the employment concentration measured by the HHI. That is the main reason why wage declines associated with higher levels of concentration are viewed as supportive evidence of anticompetitive behavior in labor markets \citep{card_who_2022}.   

My analysis so far has not distinguished the M\&A events by their predicted impact on market HHI scores, and their effect on employment concentration was measured to be positive, although imprecise, point estimates (Panel \ref{fig:g_mod_w_nyt_llm_secc3_ohhi_first_tradable_0}). In this section, I report the effects from two different types of mergers, namely \emph{out-of-market} and \emph{high-impact} M\&As. Out-of market mergers are consolidation events where the merging firms were not simultaneously active in the same labor market before the event date. In terms of employment concentration, these are the events with no predicted change in HHI, where the predicted change is the difference between a simulated measure of HHI where the two or more merging firms are considered as one single employer, and the employment HHI actually observed one year before the merger. In case the merging parties operate in the same labor market, their merger has a positive predicted change in HHI, as the sum of their employment shares is greater than any of their individual shares in the year before their merger, making the local simulated HHI greater than the one observed in the data. I label as \emph{high-mpact} the mergers at the 85th percentile, or above that, of the distribution of predicted change in HHI \footnote{One year post-merger, events at the 80th percentile and above cause a change in HHI of 253.48 points (SE=65.61), equivalent to the combination of two equally sized employers with an 11.25\% employment share each. At the 85th percentile, this effect rises to 430.70 points (SE=112.35), analogous to the merger of two  14.67\%-share partners, making it thus more suitable to test the effects of a change in HHI induced by the merger against out-of-market M\&As. Moving the threshold higher up in the predicted change in HHI distribution, however, reduces the number of contributing markets from which to infer the effects, which compromises the statistical power and feasibility of the analysis.}. The relevance of out-of-market M\&As lies in the fact that they do not mechanically induce any changes in concentration\footnote{As it soon will be shown, out-of-market M\&As do not induce changes in distribution post-treatment either.}, and thus, their effects are expected to be unrelated to market-wide anticompetitive behavior resulting from any increases in concentration. Diversely, effects from high-impact M\&As can be indicative of anticompetitive behavior related to the increase in concentration that they elicit. Table \ref{tab:dhhi_perc} shows that, out of all mergers and acquisitions in the sample of tradable industries, the median predicted change in HHI is 0, while the 85th percentile has a predicted change of 5.53 points – an event analogous to the merger of two equally sized employers with a 1.67\% share of employment each. In Table \ref{tab:om_p85_mkts} I report summary statistics of the markets used in the estimation of the effects of the two types of events. The markets in the out-of-market pool have similar earnings to those in the high-impact's, a little over 4,000 BRL, and they also depart from a close average number of firms, 81.16 and 79.83, respectively.

\begin{table}[]
\caption{\label{tab:dhhi_perc}Distribution of the Predicted change in HHI}
\vspace{-.5cm}
\centering
\adjustbox{max width=\textwidth}{%
    \begin{threeparttable}
\begin{tabular}{llllllllll}
 &                                                      &  &                                   &                                   &                                   &                                   &                                   &                                   &  \\ \hline
 & \multicolumn{1}{r}{\textbf{Percentile}}              &  & \multicolumn{1}{c}{\textbf{50th}} & \multicolumn{1}{c}{\textbf{75th}} & \multicolumn{1}{c}{\textbf{80th}} & \multicolumn{1}{c}{\textbf{85th}} & \multicolumn{1}{c}{\textbf{90th}} & \multicolumn{1}{c}{\textbf{95th}} &  \\
 & \multicolumn{1}{r}{\textbf{Predicted Change in HHI}} &  & \multicolumn{1}{r}{0.00}          & \multicolumn{1}{r}{0.10}          & \multicolumn{1}{r}{0.87}          & \multicolumn{1}{r}{5.53}          & \multicolumn{1}{r}{42.55}         & \multicolumn{1}{r}{333.11}        &  \\ \hline
 &                                                      &  &                                   &                                   &                                   &                                   &                                   &                                   & 
\end{tabular}
\begin{tablenotes}[para,flushleft]
   \vspace*{-.4cm}
   \footnotesize
   \item\textit{Notes:} The predicted change in HHI is the difference between the simulated HHI and the observed HHI one year before the M\&A takes place. The simulated HHI is obtained by considering the merging firms as one single employer in the year before the merger, effectively summing their employment shares in that year. 
   \end{tablenotes}
   \end{threeparttable}}
\end{table}

\begin{table}[]
\centering
\caption{\label{tab:om_p85_mkts}Summary Statistics of Markets in Out-of-Market and High-Impact Events}
\vspace{-12pt}
\adjustbox{max width=\textwidth}{%
\begin{threeparttable}[m]

\begin{tabular}{lrlrrlrr}
 & \multicolumn{1}{l}{}          &  & \multicolumn{1}{l}{}     & \multicolumn{1}{l}{}          &                      & \multicolumn{1}{l}{}     & \multicolumn{1}{l}{}          \\ \hline
 & \multicolumn{1}{l}{}          &  & \multicolumn{2}{c}{Out-of-Market}                        & \multicolumn{1}{c}{} & \multicolumn{2}{c}{High-Impact}                          \\ \cline{4-5} \cline{7-8} 
 & \multicolumn{1}{l}{}          &  & \multicolumn{1}{c}{Mean} & \multicolumn{1}{c}{Std. Dev.} &                      & \multicolumn{1}{c}{Mean} & \multicolumn{1}{c}{Std. Dev.} \\ \hline
 & \multicolumn{1}{l}{}          &  & \multicolumn{1}{l}{}     & \multicolumn{1}{l}{}          &                      & \multicolumn{1}{l}{}     & \multicolumn{1}{l}{}          \\
 & Market Earnings \small{(2010 BRL)}         &  & 4,066.80                 & 1,917.33                      &                      & 4,024.70                 & 1,477.92                      \\
 & Workers                       &  & 2,060.02                 & 3,055.59                      &                      & 2,342.88                 & 3,058.78                      \\
 & Workers in Merging Firms (\%) &  & 38.11                    & 40.50                         &                      & 46.13                    & 32.33                         \\
 & Firms                         &  & 81.16                    & 213.94                        &                      & 79.83                    & 169.01                        \\
 & HHI                           &  & 2,836.08                 & 2,604.09                      &                      & 2,242.50                 & 2,084.95                      \\
 & \multicolumn{1}{l}{}          &  & \multicolumn{1}{l}{}     & \multicolumn{1}{l}{}          &                      & \multicolumn{1}{l}{}     & \multicolumn{1}{l}{}          \\
 & Market-Year Observations      &  & \multicolumn{2}{c}{9,581}                                & \multicolumn{1}{r}{} & \multicolumn{2}{c}{2,886}                                \\ \hline
\end{tabular}
\begin{tablenotes}[para,flushleft]
   %\vspace*{-5pt}
   \footnotesize
   \item\textit{Notes:} Summary statistics of local labor market characteristics. Local labor markets are defined by pairs of commuting zone and 3-digit industry sector codes. Annual earnings were adjusted by inflation, with the base year in 2010. Out-of-Market M\&As are the ones with no predicted increase in HHI before the consolidation date. High-Impact M\&As are the ones in the top 15\% of predicted change in HHI. All means are computed one year before the M\&A event.      
   \end{tablenotes}
   \end{threeparttable}}
\end{table}

\begin{figure}[htp]
\begin{adjustwidth}{-1.5cm}{-1.5cm}
    \centering
    \caption{Out-of-Market and High-Impact M\&As}
    \begin{subfigure}[t]{0.49\linewidth}
        \centering
        \caption{Employment HHI}
        \footnotesize \emph{Out-of-Market M\&A}
        \includegraphics[width=\textwidth]{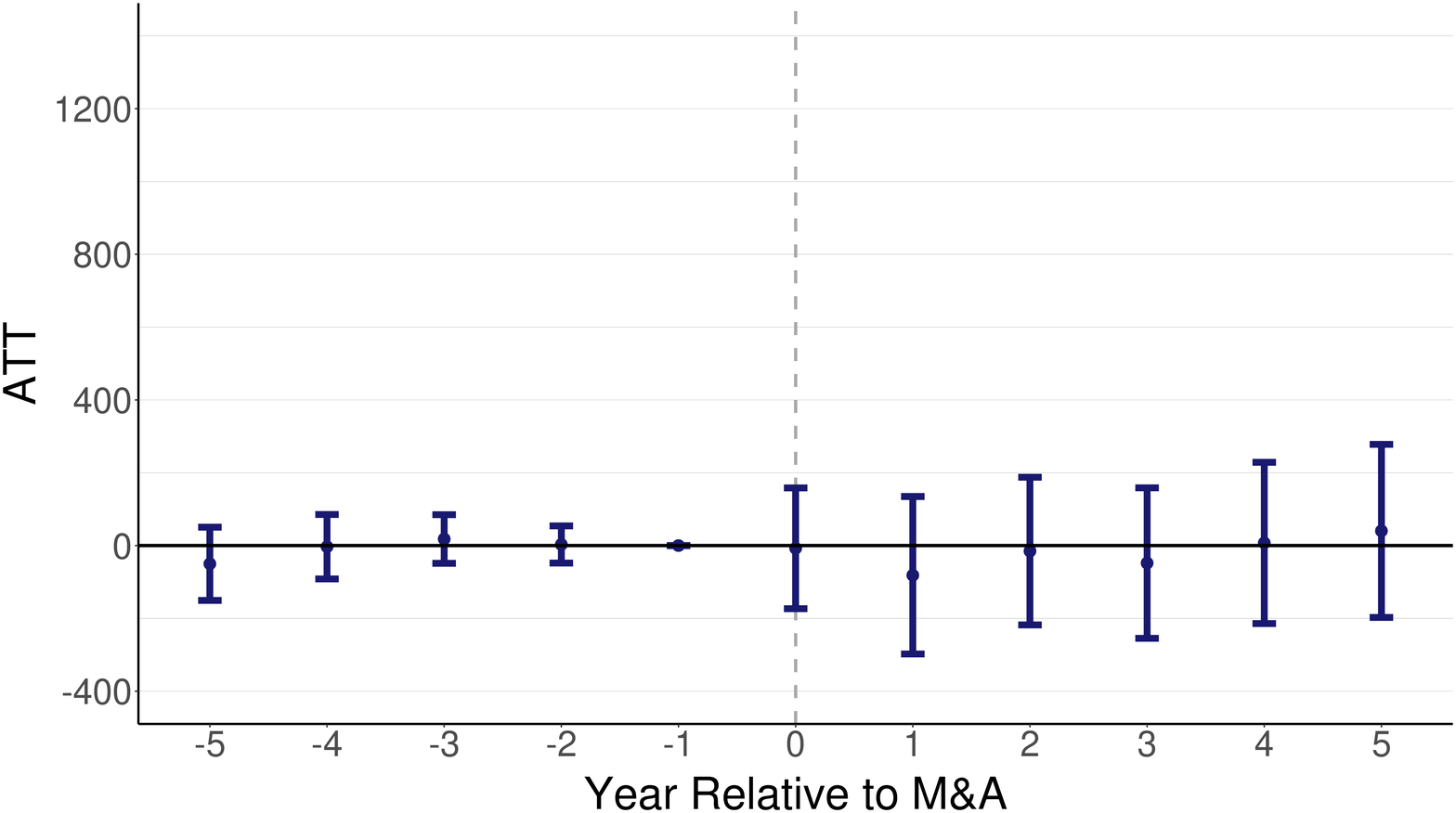}
        \label{fig:g_mod_w_nyt_llm_secc3_ohhi_first_trade_om_0}
    \end{subfigure}
    \hfill
    \begin{subfigure}[t]{0.49\linewidth}
        \centering
        \caption{Employment HHI}
        \footnotesize \emph{High-Impact M\&A}
        \includegraphics[width=\textwidth]{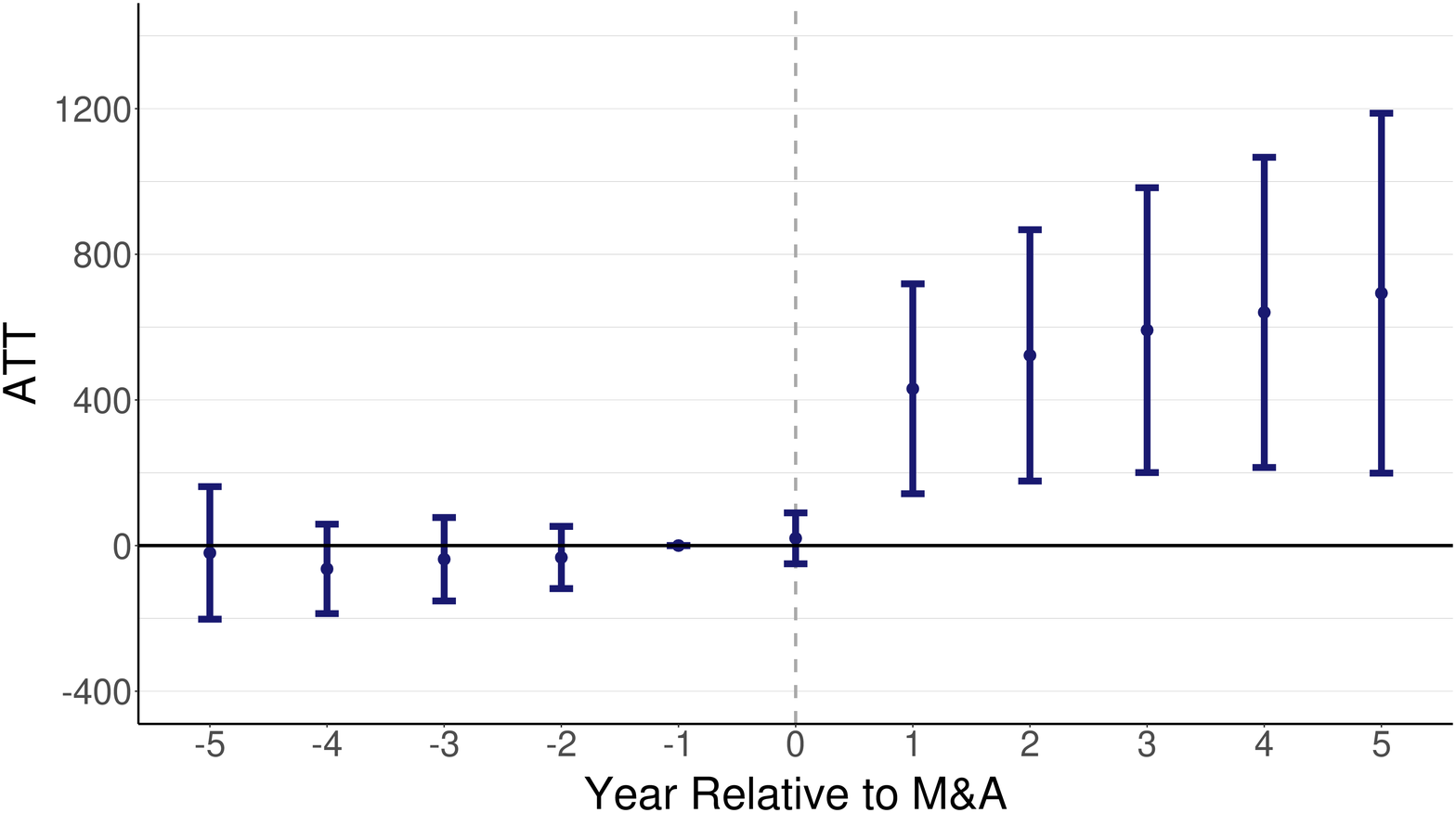}\label{fig:g_mod_w_nyt_llm_secc3_ohhi_first_trade_p85_0}
    \end{subfigure}
    \vspace{0.4cm}
    \begin{subfigure}[t]{0.49\linewidth}
        \centering
        \caption{Workers Earnings}
        \footnotesize \emph{Within M\&A Firms}
        \includegraphics[width=\textwidth]{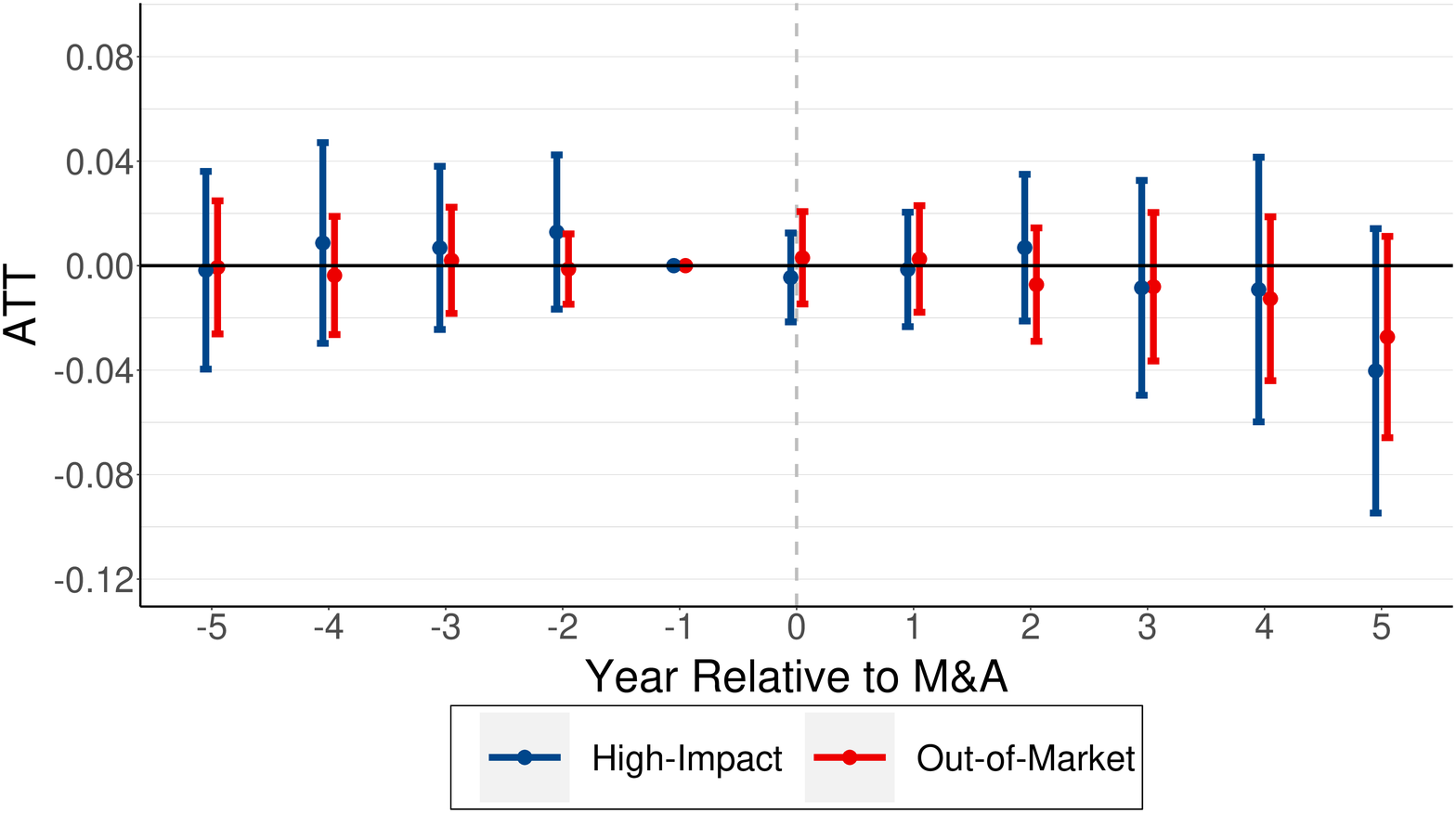}
        \label{fig:g_mod_w_nyt_llm_secc3_fe_spill0_tradable_first_trade_om_0}
    \end{subfigure}
    \hfill
    \begin{subfigure}[t]{0.49\linewidth}
        \centering
        \caption{Workers Earnings}
        \footnotesize \emph{Spillover}
        \includegraphics[width=\textwidth]{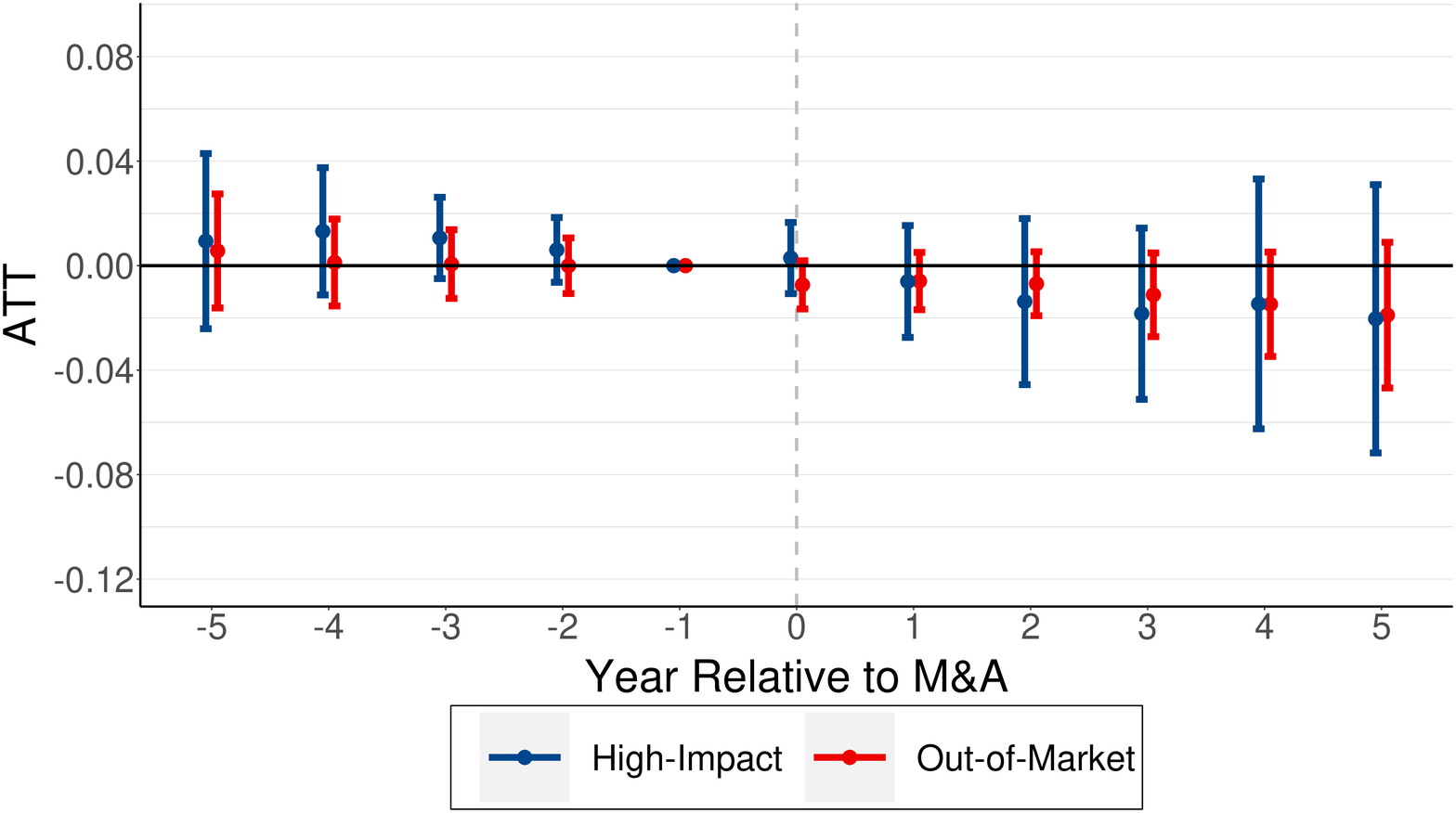}
        \label{fig:g_mod_w_nyt_llm_secc3_fe_spill1_tradable_first_trade_om_0}
    \end{subfigure}                                                                         
    \vspace{0.4cm}
    \begin{subfigure}[t]{0.49\linewidth}
        \centering
        \caption{Log Employment}
        \footnotesize \emph{Within M\&A Firms}
        \includegraphics[width=\textwidth]{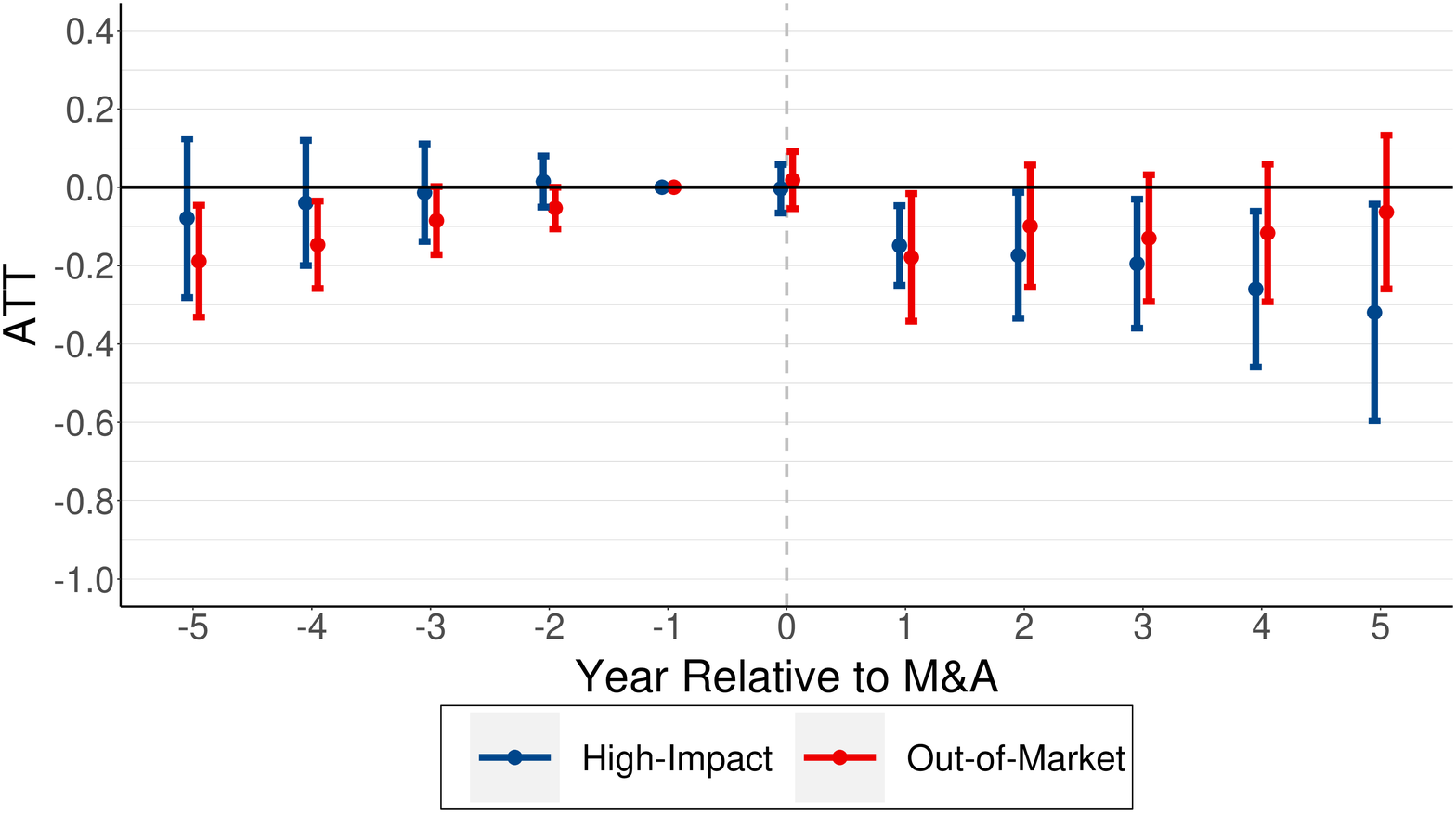}
        \label{fig:g_mod_w_nyt_llm_secc3_ur_lqt_spill0_first_trade_om_0}
    \end{subfigure}
    \hfill
    \begin{subfigure}[t]{0.49\linewidth}
        \centering
        \caption{Log Employment}
        \footnotesize \emph{Spillover}
        \includegraphics[width=\textwidth]{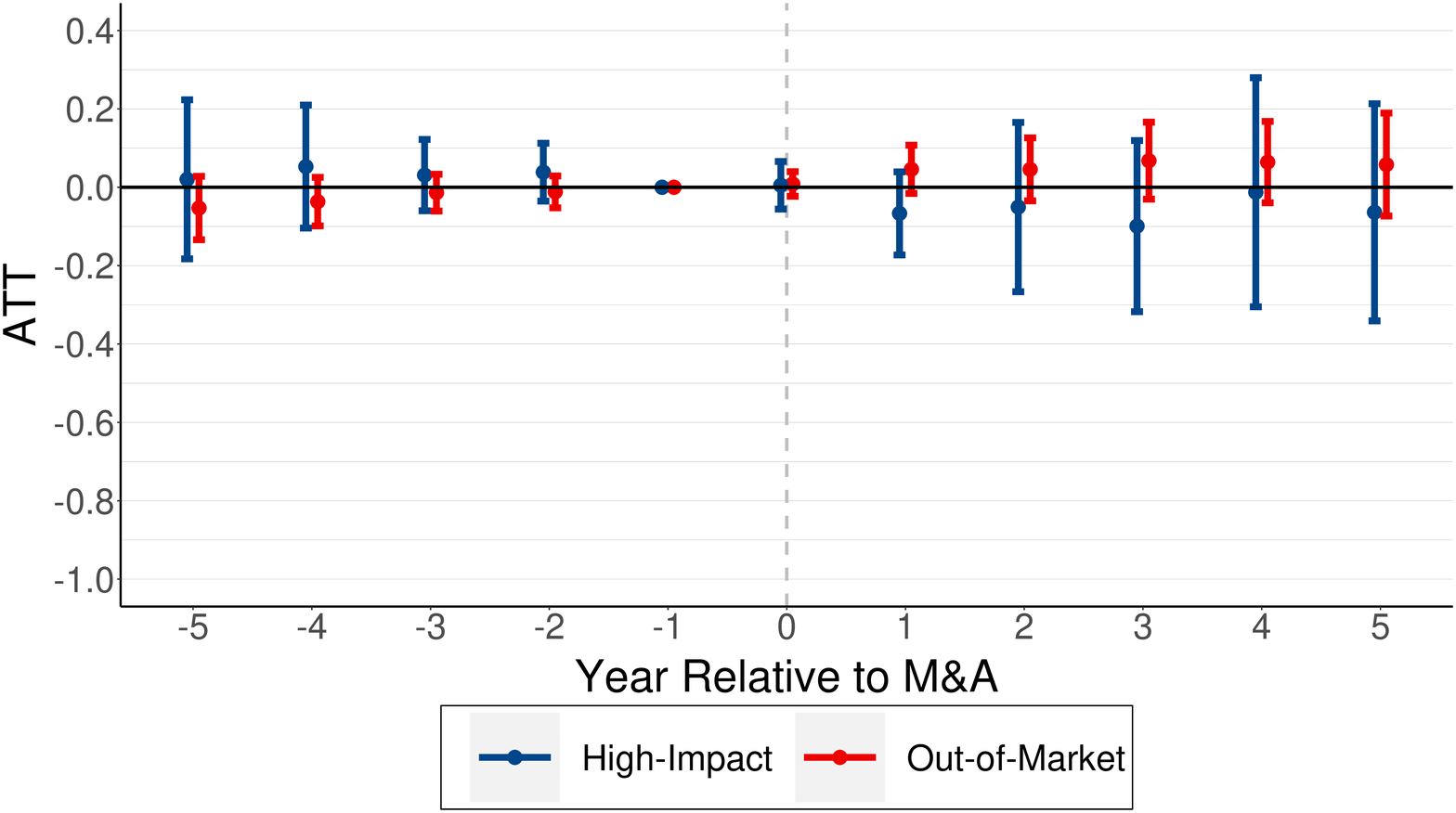}
        \label{fig:g_mod_w_nyt_llm_secc3_ur_lqt_spill1_first_trade_om_0}
    \end{subfigure}
    \label{fig:g_om_hi}
\end{adjustwidth}
\vspace{-.5cm}
    
    \scriptsize 
    \textit{Note:} 95\% confidence intervals with standard errors clustered at the market level. The year $t=0$ marks the first tradable M\&A in a local labor market, defined by pairs of commuting zone and 3-digit level industry code. Markets are weighted by their relative size in each year.
\end{figure}

In Figure \ref{fig:g_om_hi}, the first two panels confirm that out-of-market M\&As and high-impact M\&As have distinctive post-treatment concentration dynamics. As expected, high-impact M\&As increase local employment concentration – the two-year treatment exposure point estimate shows an increase in HHI of 522.48 points (SE=118.64), an increase analogous to one generated by a merger of two equally sized employers with a 16.16\% share of local employment each. Post-treatment, I cannot reject the null effects of out-of-market M\&As on local market concentration. In Panels \ref{fig:g_mod_w_nyt_llm_secc3_fe_spill0_tradable_first_trade_om_0} and \ref{fig:g_mod_w_nyt_llm_secc3_fe_spill1_tradable_first_trade_om_0}, I observe similar point estimates from both types of mergers, and in both contexts, that of merged firms and spillovers. Within merged firms, earnings seem to decline especially after the second year of exposure to treatment, while a downward trajectory is noticeable from the start in spillover firms.  As for the employment outcome, the event studies in panels \ref{fig:g_mod_w_nyt_llm_secc3_ur_lqt_spill0_first_trade_om_0} and \ref{fig:g_mod_w_nyt_llm_secc3_ur_lqt_spill1_first_trade_om_0} indicate a different pattern of effects between the two types of mergers, with negative estimates in the case of high-impact M\&As both within merged firms and spillovers\footnote{The lead estimates of out-of-market M\&A employment effects in Panel \ref{fig:g_mod_w_nyt_llm_secc3_ur_lqt_spill0_first_trade_om_0} show a potential positive linear trend in the employment of merged firms relative to their not-yet-treated counterparts in other markets. In such cases, the negative post-treatment estimates can be interpreted as a reverse in their employment growth trajectory. Either way, the qualitative conclusion remains the same and is consistent with the case of all mergers in Section \ref{sec:res_all}, employment in merged firms would have been higher if not for the merger.}. For out-of-market M\&As, while treatment effects are negative within the merged firms, they are positively estimated among spillover employers.  

\begin{figure}[htp]
\begin{adjustwidth}{-0cm}{-0cm}
    \centering
    \caption{Overall Effects by Type of M\&A}
    \begin{subfigure}[t]{0.45\linewidth}
        \centering
        \caption{Workers Earnings}
        \includegraphics[width=\textwidth]{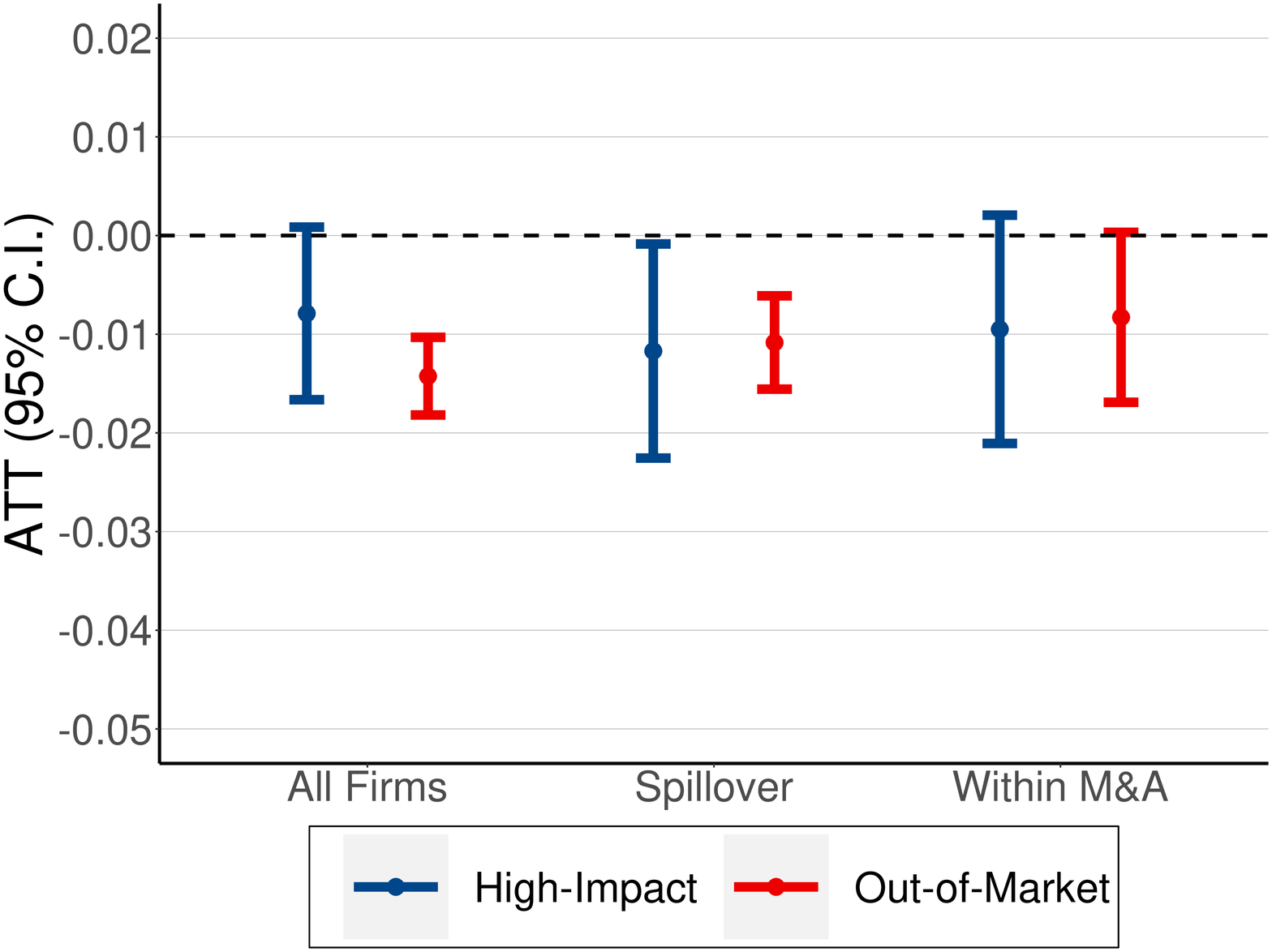}
        \label{fig:ov_earn_hi}
    \end{subfigure}
    \hfill
    \begin{subfigure}[t]{0.45\linewidth}
        \centering
        \caption{Log Employment}
        \includegraphics[width=\textwidth]{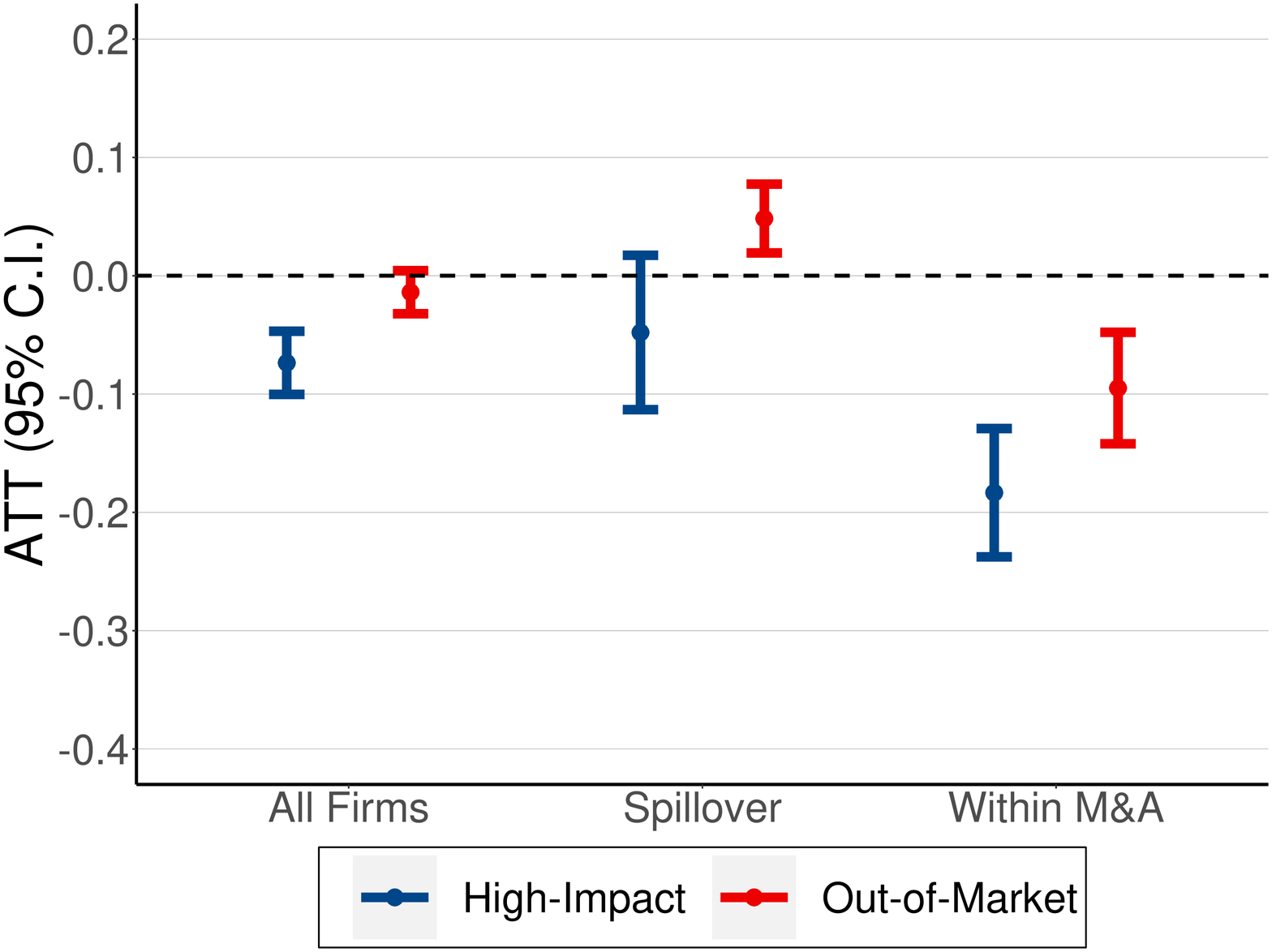}
        \label{fig:ov_emp_hi}
    \end{subfigure}
    \label{fig:ov_hi}    
\end{adjustwidth}
\vspace{.25cm}
    
    \scriptsize 
    \textit{Note:} The graph shows the ATT for all five years after treatment. 95\% confidence intervals with standard errors clustered at the market level. Markets are defined by pairs of commuting zone and 3-digit level industry code. Markets are weighted by their relative size each year. Out-of-market M\&As are mergers with no predicited change in market HHI, while high-impact are the ones in the top 15\% of the predicted change in HHI distribution.   
    
\end{figure}

The earnings and employment effects of the two types of consolidation events are summarized in Figure \ref{fig:ov_hi}. The overall estimates for the five-year post-treatment show a similarity in the magnitude of earnings decline within M\&A firms for both out-of-market and high-impact mergers, -0.0083 (SE=0.0084) and -0.0095 (SE=0.0108) respectively. In the case of spillover firms, the estimates are -0.0108 (SE=0.0051) for out-of-market mergers and -0.0117 (SE=0.011) for high-impact. This finding is at odds with studies of mergers in the context of the U.S. labor market in two aspects. First, both in the specific case of hospital consolidation \citep{prager_employer_2021}, and in the more general context of multiple industries \citep{arnold_mergers_2022}, mergers with little to no change in concentration do not have a significant impact on earnings. Here, I find a significant earnings decline after out-of-market mergers in spillover firms and, although less precisely estimated, within merged firms too. Even in the absence of changes to local concentration, mergers impart a significant decrease in earnings. Second, looking at the case of spillover firms, the effects from the two types of events are similar in magnitude, indicating a decline of 1.1\% in earnings. This similarity is surprising in view of the result that larger increases in concentration are followed by larger wage declines, both theoretically and empirically. Assuming no changes to labor supply elasticity and productivity in the first five years after the merger, a likely scenario for the case of firms not involved in the merger, the theory of oligopsony à la Cournot would have predicted a more negative wage effect for the case of high-impact events. This finding also shows that, at least in the context of Brazilian labor markets, it should not be assumed that mergers only affect wages and employment, via the concentration channel only, an assumption that has been used before in instrumental variable estimations of the relationship between wages and employment concentration \citep{benmelech_strong_2022, arnold_mergers_2022}. Without distinguishing between spillover and merged firms, the \emph{All Firms} column in Panel \ref{fig:ov_earn_hi} confirms that out-of-market mergers have a significant negative effect on earnings, estimated at -0.0143 (SE=0.0045). 

\begin{figure}[htp]
\begin{adjustwidth}{-0cm}{-0cm}
    \centering
    \caption{\label{fig:g_lfirms} Employers}
    \begin{subfigure}[t]{0.75\linewidth}
        \centering
        \includegraphics[width=\textwidth]{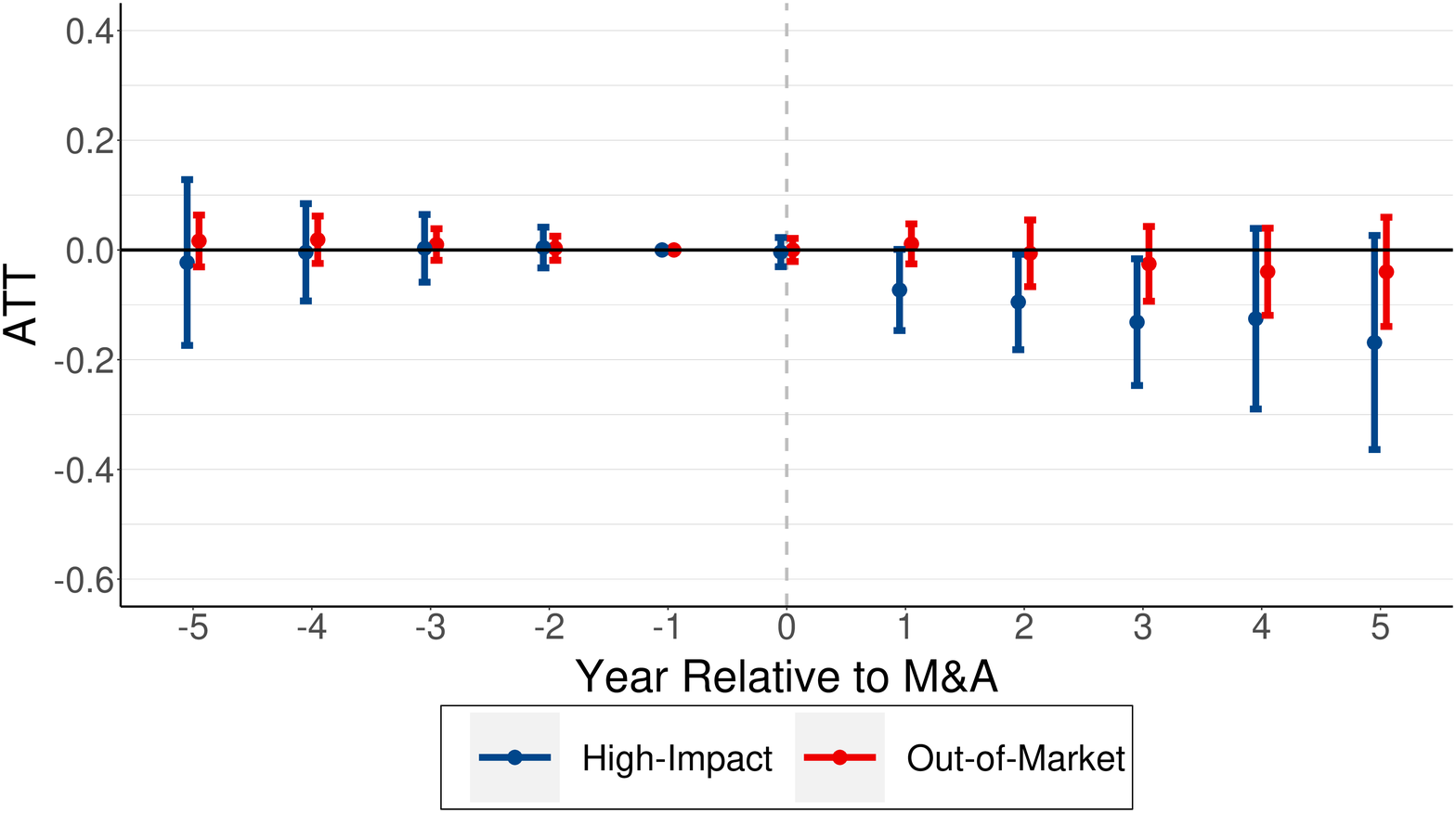}
    \end{subfigure}   
\end{adjustwidth}
\vspace{-0cm}
    
    \scriptsize 
    \textit{Note:} ATT of the M\&As on the log number of employers in a local labor market. 95\% confidence intervals with standard errors clustered at the market level. Markets are defined by pairs of commuting zone and 3-digit level industry code. Markets are weighted by their relative size each year. 
    
\end{figure}

Panel \ref{fig:ov_emp_hi} shows the overall employment effects. Here, the out-of-market mergers have a similar result to the one found in the case of all mergers presented in Section \ref{sec:res_all} - while a negative employment adjustment is observed in merged firms, spillover firms grow after the merger of their competitors, although at a rate that does not compensate for the separations in merged firms, as the \emph{All Firms} column in Panel \ref{fig:ov_emp_hi} indicates. For the case of high-impact mergers, the employment effect is negative across all firms, notwithstanding the estimates are less precisely estimated in spillover firms. In principle, this is an expected result. Contrary to out-of-market events, the high-impact mergers necessarily reduce the number of employers in the market, canceling to some extent the possibility of workers reallocating within the same labor market. Figure \ref{fig:g_lfirms} shows that the higher concentration observed after high-impact mergers may not only stem from a change in the distribution of workers among big and small firms, but also from the reduction in the number of employers altogether. The number of employers is close to 9.05\% lower in markets that witness a high-impact merger two years after the event, while a null effect cannot be rejected in case of out-of-market mergers. Another reason why employment may decline more sharply in high-impact mergers is that the merging entities might have more redundancy among their workers once under the same ownership and management, while the same level of overlap is likely not achievable in out-of-market consolidations. 

%--------------------------------------------------------

\section{\label{sec:rob}Robustness to Anticipation}
Mergers and acquisitions may take a long time to conclude, and both workers and firms can respond to the news of the event before its official date reported in administrative records. Thus, it is important to check the robustness of the results to the possibility of treatment anticipation. Effectively, this means that some not-yet-treated markets that contribute to the control pool under the assumption of no treatment anticipation are not appropriate counterfactuals anymore, specially if firms engage in pre-merger adjustments ahead of the deal taking place. To take this possibility into account, I modify the Limited Treatment Anticipation Assumption in Section \ref{ass:anticipation} by changing the value of the parameter $\zeta$ from 0 to 1. This implies that for markets treated in year $g$, their average treatment effect in any year $t$ will be based upon the not-yet-treated markets by year $t+2$, and not $t+1$ as before\footnote{It is worth noticing that this adjustment is different from simply moving the normalization period in two-way fixed effects specifications from the more commonly reported lead $t-1$ to $t-2$, in view of the fact that it effectively changes the pool of units in the control group \citep{callaway_difference--differences_2021}.}. 

\begin{figure}[htp]
\begin{adjustwidth}{-0cm}{-0cm}
    \centering
    \caption{Overall ATT with Treatment Anticipation of One Year}
    \begin{subfigure}[t]{0.45\linewidth}
        \centering
        \caption{Workers Earnings}
        \includegraphics[width=\textwidth]{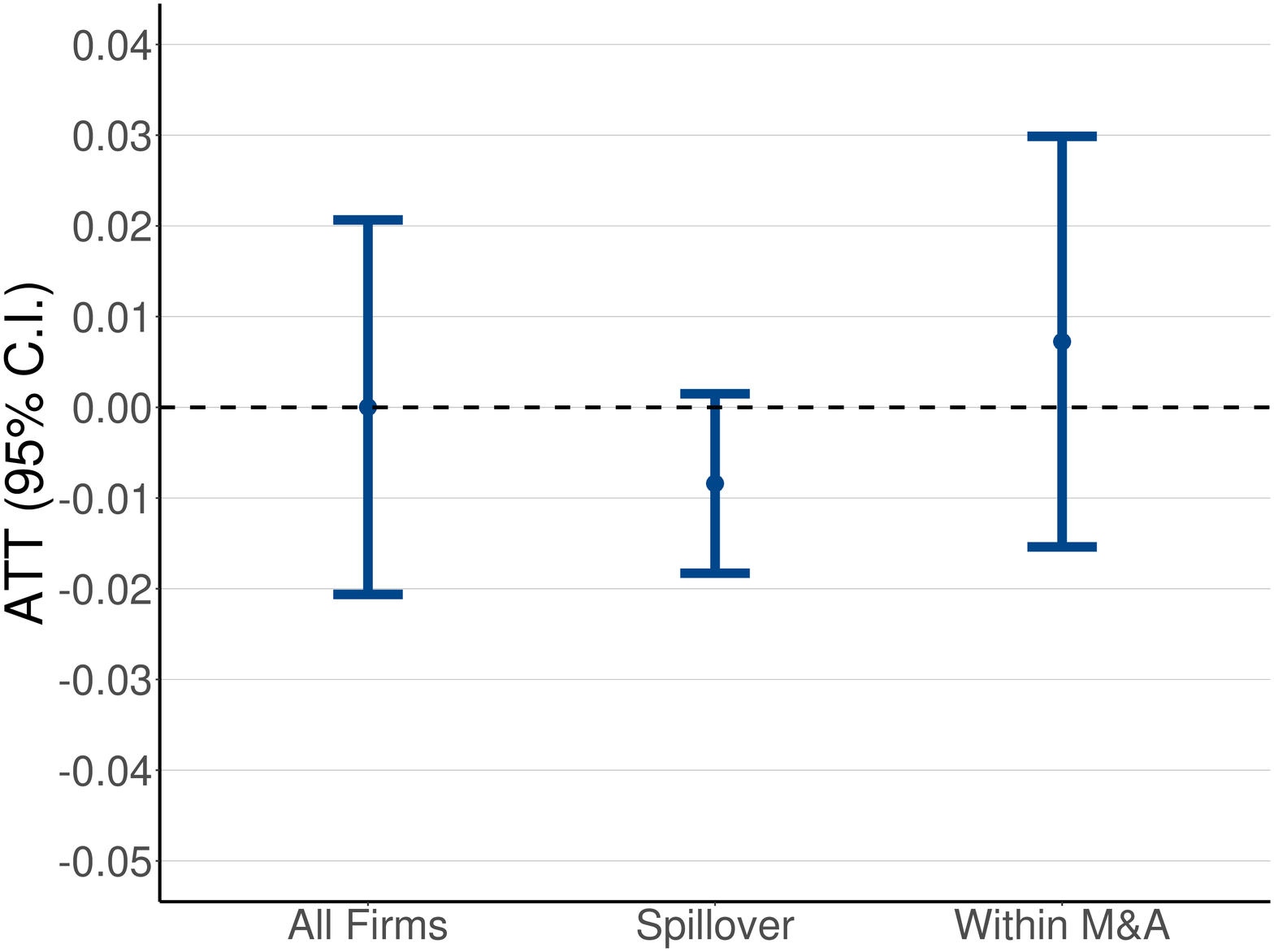}
        \label{fig:ov_earn_all_ant1}
    \end{subfigure}
    \hfill
    \begin{subfigure}[t]{0.45\linewidth}
        \centering
        \caption{Log Employment}
        \includegraphics[width=\textwidth]{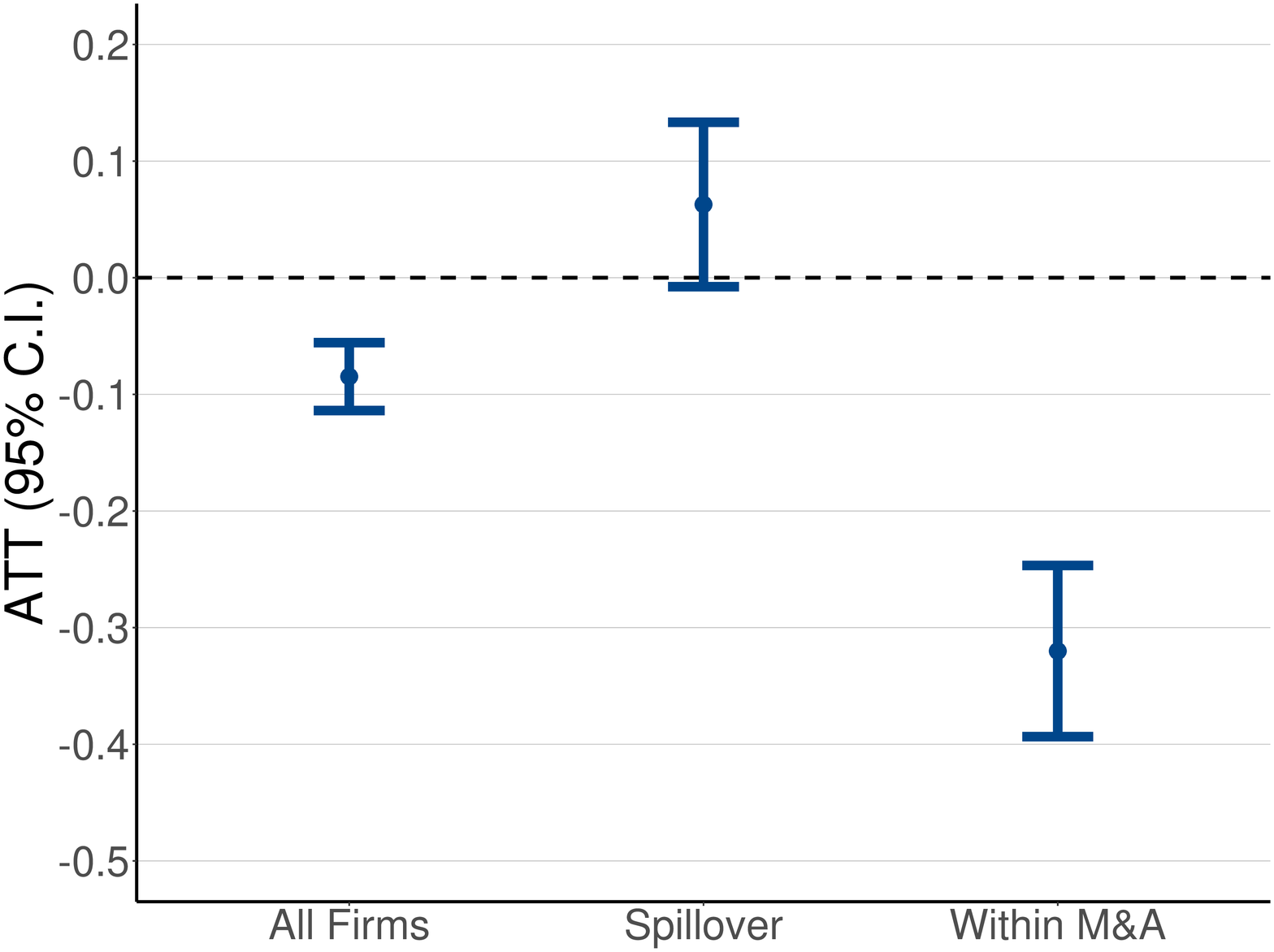}
        \label{fig:ov_emp_all_ant1}
    \end{subfigure}
    \label{fig:ov_all_ant1}    
\end{adjustwidth}
\vspace{.25cm}
    
    \scriptsize 
    \textit{Note:} The graph shows the ATT for all five years after treatment, under the assumption of treatment anticipation of one year. 95\% confidence intervals with standard errors clustered at the market level. Markets are defined by pairs of commuting zone and 3-digit level industry code. Markets are weighted by their relative size each year.   
    
\end{figure}

\begin{figure}[htp]
\begin{adjustwidth}{-0cm}{-0cm}
    \centering
    \caption{ATT with Treatment Anticipation of One Year by Type of Merger}
    \begin{subfigure}[t]{0.45\linewidth}
        \centering
        \caption{Workers Earnings}
        \includegraphics[width=\textwidth]{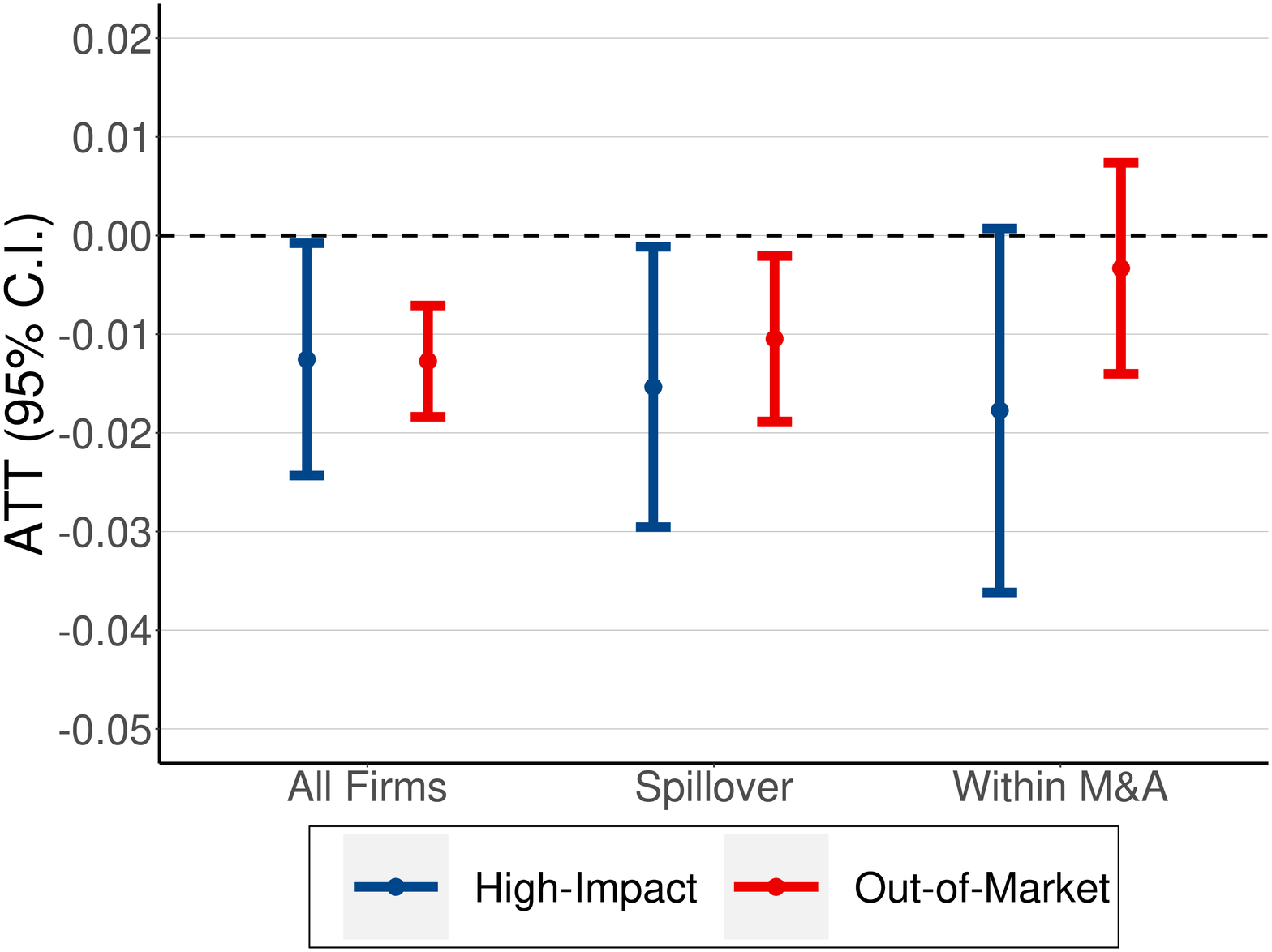}
        \label{fig:ov_earn_ant1}
    \end{subfigure}
    \hfill
    \begin{subfigure}[t]{0.45\linewidth}
        \centering
        \caption{Log Employment}
        \includegraphics[width=\textwidth]{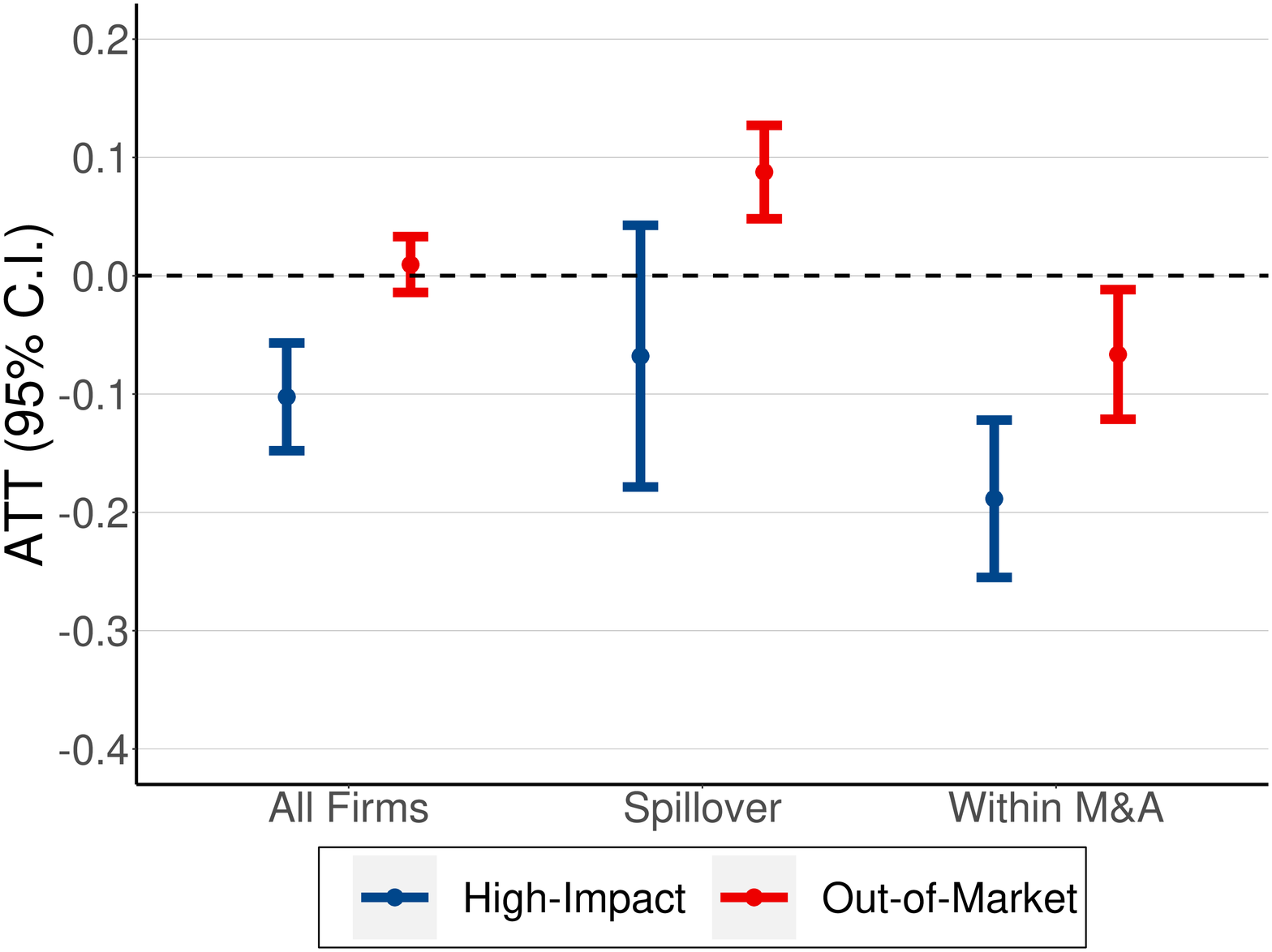}
        \label{fig:ov_emp_ant1}
    \end{subfigure}
    \label{fig:ov_ant1}    
\end{adjustwidth}
\vspace{.25cm}
    
    \scriptsize 
    \textit{Note:} The graph shows the ATT for all five years after treatment, under the assumption of treatment anticipation of one year. 95\% confidence intervals with standard errors clustered at the market level. Markets are defined by pairs of commuting zone and 3-digit level industry code. Markets are weighted by their relative size each year. Out-of-market M\&As are mergers with no predicited change in market HHI, while high-impact are the ones in the top 15\% of the predicted change in HHI distribution.   
    
\end{figure}

I present the overall 5-year ATT under the one-year treatment anticipation assumption in Figures \ref{fig:ov_all_ant1} and \ref{fig:ov_ant1}. When pooling all mergers together, the lesson from the treatment anticipation case is similar to the findings from the no treatment anticipation (Figure \ref{fig:g_ov_all}) - (i) workers earnings are lower in spillover firms, although not as precisely estimated as before, at the same time that these firms grow in size, and (ii) merging firms primarily drive the negative employment adjustment observed in the whole market, while the null effect on their workers earnings cannot be rejected. The comparison between out-of-market and high-impact events with treatment anticipation is presented in Figure \ref{fig:ov_ant1}. Here, I do not find a conclusive evidence that refutes the previous finding of similar earnings effects between the two types of events among spillover firms, which shows that even under the treatment anticipation assumption, the increase in concentration does not generate clearly distinguishable earnings effects outside the merging firms. Within merged firms, the seemingly more negative wage and employment effects of high impact events may originate from a higher degree of job titles overlap between the merging firms when they already belong to the same labor market.          

%--------------------------------------------------------
\section{\label{sec:discuss}Discussion of Results}

To summarize, mergers and acquisitions in the context of the Brazilian labor markets are shown to have significant negative employment effects at local labor market level. The post-treatment dynamics also shows increases in concentration and declines in worker earnings, although not significant. By splitting the sample between the firms that participate in the merger and all other firms in the same market, I show that the two types of firms have diverse responses on their employment and earnings margins. The negative employment effects are found primarily within the merged firms, while spillover firms show a tendency to increase in the years after the event, but their increase is not large enough to offset the reduction in size of merging firms. When it comes to earnings, I cannot reject the null effect hypothesis from the merged firms' sample, but earnings are significantly lower in spillover firms. Most mergers and acquisitions have little to no impact on concentration. It is only in the top 15\% of the distribution of \emph{a priori} increases to HHI that I find noticeable employment concentration changes. The comparison of out-of-market and high-impact mergers reveals seemingly indistinguishable earnings effects in spillover and merging firms. The out-of-market employment effects follow the overall pattern found before, merging firms get smaller and spillover firms grow in size. These findings are robust to the possibility of treatment anticipation, and are likely not related to changes in the composition of the labor force (given the construction of the earnings variable) and changes in product market power (given the restriction to tradable industry sectors only). 

The concentration channel connecting mergers and negative wage effects found in \cite{prager_employer_2021} and \cite{arnold_mergers_2022} seems to be thus absent in the context of Brazilian labor markets. In addition, I find negative wage effects that affect other firms in the labor market even in the case of merger activity not followed by increases in concentration. I do not empirically find a confirmation of the connection between increases in concetration and sharper wage declines from oligopsony models of the labor market \citep{boal_monopsony_1997, azar_measuring_2019}. But if not changes in concentration, what could be driving the observed decline in earnings of firms not related to the M\&As and the market-wide decline in employment?

\subsection{M\&A's Synergies, Managerial Practices, and Within-Market Dynamics}
\begin{figure}[htp]
    \centering
    \caption{New Hires' Earnings in Spillover Firms After an Out-of-Market Merger}
    \begin{subfigure}[C]{.7\textwidth}
        \centering
        \includegraphics[width=\textwidth]{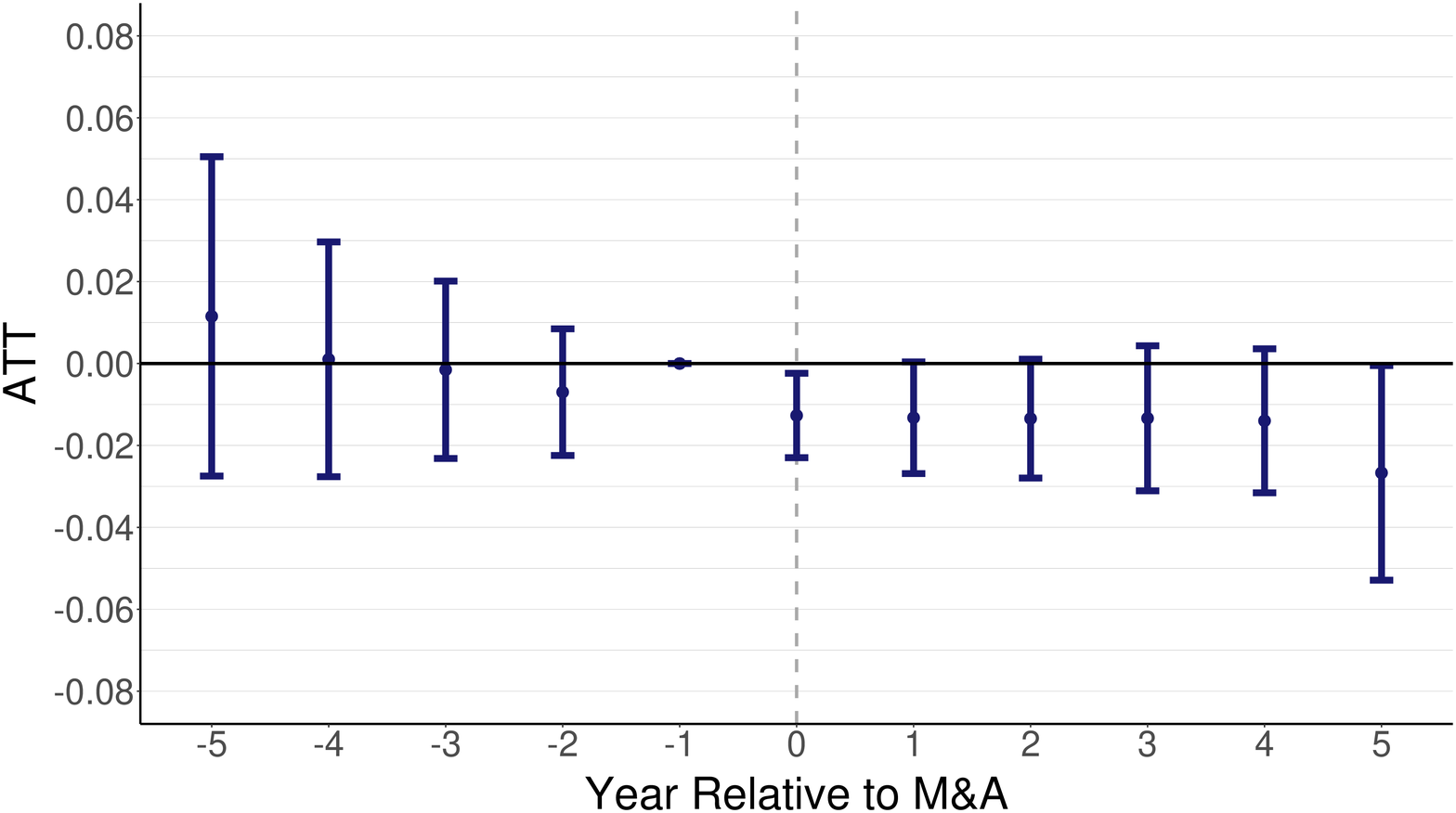}
    \end{subfigure}
    \label{fig:nh_spill_om}
    \vspace{.5cm}
    \begin{minipage}{.7\textwidth}
    \footnotesize  
    Note: Local labor markets are defined by pairs of commuting zone and 3-digit industry code. New hires' earnings refer to the log wage adjusted from equation \ref{eq:llm_wage} estimated among the pool of new hires in spillover firms. New hires are workers with less than 12 months of tenure in the current job by December 31st of each year. Spillover firms are the employers in the labor market that do not participate in any M\&A neither as acquirers, targets, nor merging partner. An out-of-market merger is a merger with zero predicted change in local concentration measured by the HHI.               
    \end{minipage}
\end{figure}

A way to rationalize the decreases in workers' earnings and employment, even in the case of mergers that do not affect local concentration, is to admit the possibility of efficiencies created by employer consolidation. Merger proponents argue that cost-saving measures can be taken once the merging parties operate under the same ownership. While the economics profession has been skeptical of efficiency claims made by merger proponents\footnote{For a discussion on the credibility of such claims in the U.S. context, see \cite{fisher_horizontal_1987} and \cite{farrell_scale_2000}. The possibility of merger-related efficiency gains is also explicitly acknowledged by regulators. See, for example, the DOJ-FTC Horizontal Merger Guidelines, Section 4.}, the present case shows that merging firms do engage in a significant reduction of personnel, while other employers in the labor market show a tendency to grow after the merger. On the earnings margin, the results showed that workers that remain in the merging firms do not face a significant decline in earnings, while the opposite happens with spillover firms' workers. What this contrast suggests is that the adjustment towards lower employment generated by potential efficiency gains in merging firms increases the supply of labor to all the other firms in the same market. Assuming a stable curve of demand for labor in spillover firms, the increased supply of workers is accommodated at a higher equilibrium level of employment and lower earnings in these firms. Indeed, I estimate negative earnings effects among new hires in spillover firms after an out-of-market merger (Figure \ref{fig:nh_spill_om}), where the fiver year post-treatment average effect is -0.0156 (SE=0.0051).     

The question of why cost-saving measures in the form of overhead reduction are more pervasive in Brazilian merger activity if compared to the context of the U.S. labor market remains\footnote{Due to a pre-trend in their employment event study, \cite{prager_employer_2021} refrain from making an assertive conclusion about the employment effects of hospital mergers. At the same time, \cite{arnold_mergers_2022} finds negative employment effects in M\&A establishments that range from 5\% to 10\% on average depending on the predicted change in concentration, I find a sharper decrease of 23.07\% in merging firms' size in the case of all mergers.}. One possibility is that changes in ownership and management are able to collect higher gains from efficiency in emerging economies due to inadequate management practices in target firms. \cite{bloom_why_2010} presents a comparison between the productivity of firms in developing economies vis-à-vis their counterparts in richer countries. In 2005, the sales per employee in American firms was more than 3.2 times as large as that of Brazilian firms. While previous studies have elicited structural, economy-wide reasons for the productivity gap, such as developing countries' lack of infrastructure, lower human capital, and regulation, Bloom conjectures that managerial practices can also be playing an important role. Compared to higher income economies, middle and low-income countries, including Brazil, have a lower prevalence of management practices related to clear target setting, production monitoring, and proper pay incentives \citep{scur_world_2021}. To the extent that changes in ownership in Brazil can allow merging firms to reap managerial gains and dismiss excess workers in the process, this could in part explain the difference in employment effects between merging firms and other employers in the same labor market. Simultaneously, M\&As in developed countries such as the U.S. might not be able to collect the same cost related efficiency gains given their superior management practices beforehand.   

%-------------------------------------------------------
\section{\label{sec:conc}Conclusion}

What are the effects of merger and acquisition activity in the labor markets of a middle-income country? I attempt to answer this question by exploring linked employer-employee administrative records from Brazil to identify merger events, locate them in labor markets defined by pairs of industry and commuting zone, and, by means of an event study design, I estimate their impact on workers earnings, employment, and local concentration measured by the HHI. The worker-flow identification of merger events allows me to distinguish the changes in size and worker compensation, both in merging firms and all the other employers doing business in the same labor market. Overall, mergers have clear negative impacts on labor market size, while null effects on earnings and local concentration cannot be rejected. I find that the market's negative employment adjustment is exclusively concentrated in merging firms, while other employers in the same market experience a positive, although not significant, size effect, at the same time that their workers'earnings show a subtle modest trend. 

The apparent null effect of mergers in local concentration is explained by the fact that most M\&As are of the out-of-market type, i.e., either the acquirer or the merging partner were not active in that same market before, and thus the predicted change in local HHI is zero. It is only at the top 15\% of the predicted change in HHI distribution that I find a noticeable local concentration impact of M\&A events. Contrary to the previous literature findings, workers' earnings in spillover firms decline similarly irrespective of the impact on concentration from the merger event. The earnings decline in spillover firms can be rationalized by a positive shift of the labor supply curve in these firms, originated by a halt in hiring from the merging competitors. At the margin, I confirm that new hires in spillover firms earn relatively lower wages after an out-of-market event. By comparing the effects of M\&As inducing different changes in local employment concentration, this paper also adds to the empirical investigation of oligopsony models that predict lower wages in concentrated labor markets.    

The body of evidence showing the negative relationship between employment concentration and labor outcomes in developed economies has prompted the suggestion that antitrust authorities should use HHI benchmarks to flag mergers' potential anticompetitive impacts on labor markets \citep{naidu_antitrust_2018}. In contrast, by explicitly comparing out-of-market and high-impact mergers earnings' effects, I find that wage declines in spillover firms are similar in both cases. However, the finding that mergers with substantial increases to concentration are not followed by stronger wage reductions should not be taken as proof that labor markets in Brazil are hence perfectly competitive, and that the local antitrust authority should not be cognizant of mergers impacts on workers. The oligopsony theory is one way to rationalize wage markdowns, but models of job search friction, such as employer differentiation, and job ladder, can also generate firm-specific upward sloping labor supply curves independent of employment concentration\footnote{See \cite{card_who_2022} for a comparison and historic perspective on models with employer's wage setting power.}. What my result shows is that \emph{ad-hoc} thresholds of local concentration may not be as informative about the competitiveness of developing economies' labor markets as they are in developed countries. Additionally, the existence of non-compete clauses common in high-skilled occupations, and anti-poaching agreements documented among various U.S. franchisees are worth the attention of antitrust policy regardless of their connection to local market concentration \citep{balasubramanian_locked_2020, krueger_theory_2022}. The pervasiveness of such practices in the Brazilian context is still unexplored, and undoubtedly deserve the attention of future research.

%--------------------------------------------------------

\newpage
\bibliographystyle{aer}
\bibliography{library} %bibtex file name without .bib

%--------------------------------------------------------
\newpage
\appendix

\section{\label{app:data_mani} Data Handling}

\subsection{Preparing RAIS}

The \emph{Relação Anual de Informações Sociais} - RAIS version used in this paper starts in 2002 and ends in 2017. The files are made available by the Ministry of Labor in Brazil, and their transfer is conditional on a confidentiality agreement celebrated between Cornell's Labor Dynamic Institute and the Ministry. The files are hosted in a secured cluster within Cornell's BioHPC server ecosystem. RAIS' raw files are year-by-region (in some years, year-by-state) \texttt{.txt} tables. I fist stack all files within the same year, then I merge the yearly files with the commuting zone list using the city codes as the key. Job records outside the reach of commuting zones are dropped.   

For each job record, RAIS reports the status of the job in December 31st – if this variable has entry 0 it means that the work contract was terminated at some point in that year. I keep only the job records with active employment contracts on December 31st. As standard, in case the same worker has more than one employer, I keep the highest paying job. For the demographics used in the estimation of \ref{eq:llm_wage}, I use the worker's \emph{age}, \emph{age squared}, and dummy variables for \emph{female}, \emph{white}, \emph{college or higher} (codes greater than or equal to `9'), and \emph{high school} (codes `7' or `8'). RAIS is an employer-reported database, and in some occasions, a worker's history will show different colors/races, depending upon either the perception of the current employer or the worker's informed race when the job started \citep{cornwell_wage_2016}. If a worker is ever reported as \emph{non-white} (race/color codes different than `2'), I set the \emph{white} dummy to 0. 

In Section \ref{sec:res_split} I present event studies looking at the dynamic of new hires and separations within merged and spillover firms. In order to exclude spurious work contract terminations, such as transfers across establishments of the same firm or to the newly merged firm, I exclude the separations with reported reason coded with labels `30' and `31'. New hires are flagged using the tenure variable, measured in months as of December 31st of each year. Active jobs on December 31st with tenure less than or equal to 12 months are flagged as new hires. In order to avoid counting spurious admissions, similar to the case of separations, I exclude the admissions coded with types `3' and `4'.

\subsection{\label{app:top_dest}Identification of Establishments in M\&A events}

The identification of merger activity starts with the \emph{Dados Públicos de CNPJ} - DPC, released monthly by the revenue agency in Brazil. The release I used in the paper is from September 5th, 2020. The file contains approximately 42.5 million observations at establishment level. Every time an establishment is acquired or merged with others, its identifier is retired, and a new one is issued by the revenue agency. The establishment identifier, also called the establishment's CNPJ, is a hierarchical 14-digit code, where the first eight digits identify the firm, the following four digits identify the establishment, and the last two digits are used for checksums. I rely on the column describing the reason for retirement of a CNPJ to flag acquired (code `2') or merged (code `3') establishments.    

\begin{figure}[htp]
    \centering
    \caption{\footnotesize \textbf{Distribution of the Percentage of Workers Departing From Acquired or Merged Establishments to Most Common Destination Firm}}
    \includegraphics[width=0.7\textwidth]{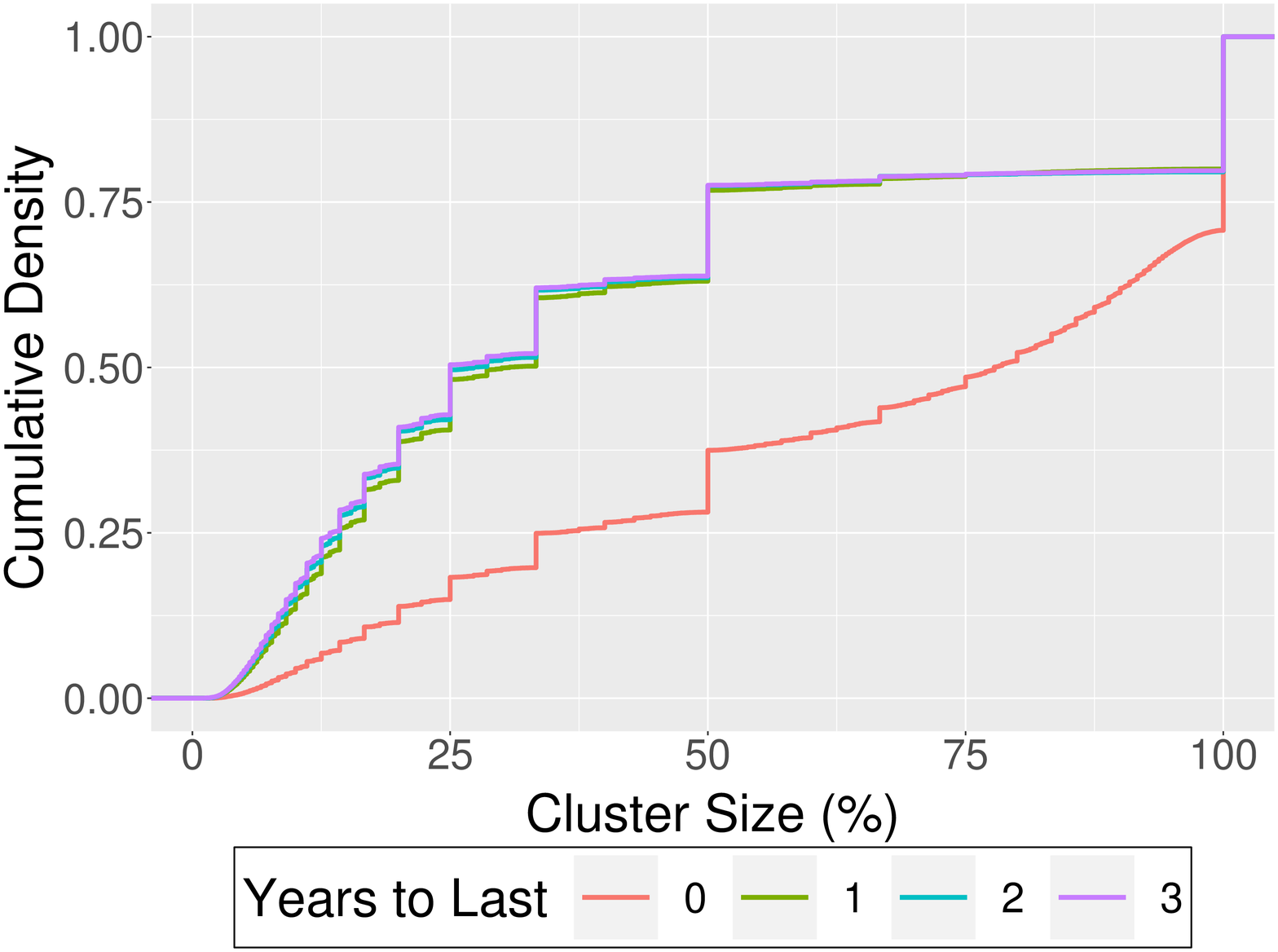}
    \label{fig:top_dest}
    \vspace{.5cm}
    \begin{minipage}{.7\textwidth}
    \small  
    Note: Consider that year $T$ is the last year of an acquired or merged establishment in RAIS, the worker data. There is a positive fraction of workers leaving acquired or merged establishments to the most common destination in years $T-1$, $T-2$, and $T-3$. In year $T$, however, the fraction of workers decamping towards the same firm is significantly larger than in previous years. The most common destination firm of workers leaving in year $T$ is flagged as the acquirer or newly merged firm. 
    \end{minipage}
\end{figure}

After flagging the acquired or merged establishments, I resort to the matched character of RAIS to keep a record of the next destination of workers who are re-employed in other firms. For each destination firm, I compute the relative size of the leaving coalition, and choose the largest coalition's next firm as the most common destination. Figure \ref{fig:top_dest} shows the distribution of worker coalition sizes departing from acquired or merged establishments towards the most common destination firm in the last few years of such establishments. In the last year of an acquired establishment, in at least 70\% of cases, more than 50\% of workers are reported to be working at the same top destination firm in the following year. In the second to last year, and before that, a coalition of less than 10\% of the acquired establishment workers can be found in the following year's most common employer. Therefore, I declare the firm that admits the most number of workers from an acquired or soon-to-be merged establishment, after the last year of observation of this establishment, as the buyer, or newly-merged firm, side of the M\&A. The identification of both acquirer and acquired allows me to compute the predicted impact on HHI of each local labor market event, or events, in a given year. The predicted HHI is then used to separate out-of-market events from those that induce higher concentration changes. 

\end{document}